\documentclass[aoas,preprint]{imsart}

\RequirePackage{amsthm,amsmath,amsfonts,amssymb}
\RequirePackage[authoryear]{natbib}
\RequirePackage[colorlinks,citecolor=blue,urlcolor=blue]{hyperref}
\RequirePackage{graphicx}

\usepackage[ruled,vlined]{algorithm2e} 
\usepackage[figuresright]{rotating}
\usepackage{float}

\usepackage{tikz} 
\usetikzlibrary{fit,positioning}
\usetikzlibrary{matrix}
\usetikzlibrary{calc}
\usetikzlibrary{arrows}
\tikzstyle{int}=[draw, fill=white!8, minimum size=2em]
\tikzstyle{init} = [pin edge={to-,thin,black}]

\begin{document}

\begin{frontmatter}

\title{Inference for stochastic kinetic models from multiple data sources for joint estimation of infection dynamics from aggregate reports and virological data}
\runtitle{Inference for SKMs from multiple data sources}


\author{\fnms{Oksana A.} \snm{Chkrebtii}\thanksref{t1,t2}\ead[label=e1]{oksana@stat.osu.edu}},
\author{\fnms{Yury E.} \snm{Garc\'ia}\thanksref{t1,t3}\ead[label=e1]{yury@cimat.mx}},
\author{\fnms{Marcos A.} \snm{Capistr\'an}\thanksref{t3}\ead[label=e1]{marcos@cimat.mx}},
\and
\author{\fnms{Daniel E.} \snm{Noyola}\thanksref{t4}\ead[label=e1]{dnoyola@uaslp.mx}}

\thankstext{t1}{Indicates equal contribution}
\thankstext{t2}{Department of Statistics, The Ohio State University, Columbus, Ohio, USA}
\thankstext{t3}{\'{A}rea de Matem\'{a}ticas B\'{a}sicas, Centro de Investigaci\'on en Matem\'aticas, Guanajuato, Gto., M\'exico}
\thankstext{t4}{Department of Microbiology, Faculty of Medicine, Universidad Aut\'onoma de San Luis Potos\'i, M\'exico}

\begin{abstract}
Before the current pandemic, influenza and respiratory syncytial virus (RSV) were the leading etiological agents of seasonal acute respiratory infections (ARI) around the world. In this setting, medical doctors typically based the diagnosis of ARI on patients' symptoms alone and did not routinely conduct virological tests necessary to identify individual viruses, limiting the ability to study the interaction between multiple pathogens and to make public health recommendations. We consider a stochastic kinetic model (SKM) for two interacting ARI pathogens circulating in a large population and an empirically-motivated background process for infections with other pathogens causing similar symptoms. An extended marginal sampling approach, based on the linear noise approximation to the SKM, integrates multiple data sources and additional model components. We infer the parameters defining the pathogens' dynamics and interaction within a Bayesian model and explore the posterior trajectories of infections for each illness based on aggregate infection reports from six epidemic seasons collected by the state health department and a subset of virological tests from a sentinel program at a general hospital in San Luis Potos\'{i}, M\'{e}xico. We interpret the results and make recommendations for future data collection strategies.
\end{abstract}


\end{frontmatter}

\section{Introduction}\label{sec:introduction}

Inference on mathematical models of epidemic dynamics has become a powerful tool in the understanding and assessment of disease outbreaks (\cite{huppert2013mathematical,siettos2013mathematical,star2010role}). Such models are predominantly stochastic, reflecting the inherently random nature of a large number of human interactions which enable infections to spread and individuals to change their infection status. The probabilities of discrete transitions from one infection state to another are defined up to a set of unknown parameters which may be inferred from observed data. The most widely used models are variations on the ``Susceptible-(Exposed-)Infected-Recovered'' (SIR/SEIR) formulation which describes the temporal evolution of the number of individuals in each infection state at a given time. A variety of strategies have been developed to incorporate process-specific demographic stochasticity in this compartmental model. For example, \cite{Dukic2012} model process stochasticity by an additive white noise process on the growth rate of the infectious population computed from states that evolve according to the deterministic compartmental dynamics described above. In a different approach, \cite{FarahEtAl2014} assume additive process noise on the infection states of a deterministic SEIR model. Another approach is taken by \cite{shrestha2011statistical} by modeling infection state counts as multinomial processes with probabilities of inclusion obtained by first solving the ODE corresponding to the compartmental model and then solving for the transition probabilities as functions of current states. Another popular class of models, considered in this paper, probabilistically models individual interactions and their outcomes directly. This approach is based on stochastic kinetic modeling (e.g., \cite{wilkinson2006stochastic}), a first-principles interpretation of SIR dynamics inspired by chemical mass-action kinetics where molecules interact and change state according to specified transition probabilities. This approach realistically captures inherent stochasticity in infection states because it naturally models individual-level transitions from one infection state to another as stochastic processes incorporating assumptions about both the disease and the interactions. However, estimation for such models poses a unique set of challenges. For a small population or one where state transitions can be observed directly, posterior inference on the SKM parameters can exploit closed-form expressions for full conditional distributions (\cite{BoysEtAl2008,ChoiRempala2011}), but even for moderately sized populations, individual interactions and transitions become increasingly difficult to simulate and marginalize over. Furthermore, since data typically consists of observed infection counts rather than individual transition times, computation of the likelihood becomes computationally infeasible. To resolve this issue, a common strategy is to construct the likelihood based on a large-population limit via diffusion approximation of the SKM (\cite{van1992stochastic}), also known as the linear noise approximation (LNA). The LNA has been a popular approach for inference for stochastic reaction networks in epidemiology and systems biology (e.g., \cite{fearnhead2014inference,fintzi2020linear,Golightly2015,GolightlyWilkinson2011,komorowski2009bayesian}) and has been successfully implemented within a Bayesian hierarchical modeling framework (\cite{Finkenstdt2013,HeyEtAl2015,WhitakerEtAl2017}). Challenges in applying this approach to realistic data scenarios include integrating multiple data types that depend on the unknown infection states, empirical modeling of disease states that are not described by the mechanistic model, and efficient posterior sampling of model parameters. The contribution of the present work is to simultaneously address these challenges to enable joint inference and retrospective assessment of the dynamics of two interacting pathogens from both aggregated infection reports and a subset of virological data collected in the state of San Luis Potos\'{i}, M\'{e}xico. The ability to utilize multiple data sources allows estimation of parameters involved in the interaction of different pathogens that cause similar symptoms. Indeed, to our knowledge our approach is the first to directly recover cross-interference parameters in a SKM for two pathogens within a Bayesian model. We additionally identify and address practical challenges of working with such models, including posterior sampling strategies that are well suited to these settings.

The results of this analysis can play an important role in epidemiological studies of the burden of influenza on morbidity and mortality at local, national, or regional levels. Most current estimates of influenza-associated morbidity and mortality do not take into account the contribution of RSV to excess mortality, due mainly to the scarcity of virological surveillance information. Our findings provide evidence that maintaining a sentinel program, by administering virological tests at a clinical referral center, allows for reliable tracking of virus circulation time at the population level at a relatively low cost. Our analysis shows that even a relatively small number of such virological tests, when administered to all the patients in a particular group, can help to provide information about patterns of pathogen circulation over time.

The article is organized as follows. The motivating application and the data are described in detail in Section~\ref{sec:data}. Section~\ref{sec:Modeling} constructs a SKM for the evolution of individual infection states of the acute respiratory pathogens influenza and RSV and formulates its behaviour in the large-population limit as a hidden Markov model. A Bayesian model is then presented relating aggregated infection counts and a subset of virological tests to the SKM for infection states of the two pathogens augmented by a background model component representing infection by other pathogens. The inferential strategy is described in Section~\ref{sec:methodology}. Section~\ref{sec:results} describes and interprets the results of our retrospective analysis for six individual epidemic seasons, using predictive residual analysis to assess model performance and to guide future modeling efforts. Finally, Section~\ref{sec:discussion} discusses the feasibility of our approach, summarizes our findings, and offers some perspectives on future work.

\section{Motivating application}\label{sec:data}

ARI are infections of the upper and lower respiratory tract caused by multiple etiological agents, most frequently, adenovirus, influenza A and B, parainfluenza, RSV, and rhinovirus. An important public health concern around the world, ARIs are responsible for substantial mortality and morbidity (\cite{thompson2003mortality,troeger2018}), mainly affecting children under five and adults over 65 years of age (\cite{kuri2006mortalidad}). Understanding of the underlying mechanisms of spread, transmission, and, importantly, cross-interference between these pathogens aids policy makers in assessing public health strategies and decision-making (\cite{huppert2013mathematical}).

Although different viruses are responsible for ARI, a substantial part of the burden of ARI in most regions has been due to influenza and RSV (\cite{chan2014robust,chaw2016burden,velasco2015superinfection}). The interaction and temporal dynamics of these pathogens are complex. Evidence suggests that influenza and RSV are seasonally related (\cite{bloom2013latitudinal,mangtani2006association}) and circulate at similar times of the year in some temperate zones (\cite{bloom2013latitudinal,velasco2015superinfection}). Because of their interaction and interference, these infections do not usually reach their epidemic peaks simultaneously (\citeauthor{Anestad1987} (\citeyear{anestad1982interference,Anestad1987}), \cite{anestad2009interference}), with peak times typically differing by less than one month (\cite{bloom2013latitudinal}). In the clinical setting it is difficult to determine which pathogen may be responsible for a patient's ARI because of their overlapping circulation times and similar symptoms. Furthermore, laboratory tests necessary for identification of the virus are not conducted in most patients (\cite{chan2014robust}).

Our analysis is based on data from epidemic seasons 2003--2004 through 2008--2009, beginning in the first week of August and ending in the last week of July of the following year. Specifically, we use data on weekly aggregated morbidity reports and virological time series obtained in the state of San Luis Potos\'{i}, M\'{e}xico. Morbidity reports consist of weekly reports of ARI that required medical attention\footnote{This category includes all cases corresponding to the International Classification of Disease, 10th review (ICD-10) codes: J00--J06, J20, J21, except J02.0 and J03.0} at all community clinics and hospitals in the state that were reported to the State Health Service Epidemiology Department. The virological data was obtained from records of the Virology Laboratory (Facultad de Medicina, Universidad Aut\'{o}noma de San Luis Potos\'{i}, San Luis Potos\'{i}, M\'{e}xico) and consisted of weekly counts of positive tests for RSV and influenza as well as tests which were negative for both pathogens. These virological tests were performed on all eligible pediatric patients (children under five years of age) admitted with lower respiratory infections (\cite{GomezVillaEtAl2008,VizcarraUgaldeEtAl2016}) to Hospital Central ``Dr. Ignacio Morones Prieto''. These virological tests are unlikely to reflect only small-scale outbreaks, since Hospital Central ``Dr. Ignacio Morones Prieto'' admits patients from all regions of the state. Though the number of virological tests is small relative to the total number of aggregated ARI reports, the systematic sampling of patients year-round, irrespective of the identification of an epidemic by any given virus, provides information on the period of circulation of influenza and RSV. As in our data, most virological tests for RSV are performed on samples from children (\cite{amini2019respiratory,MeskillEtAl2017}) and may be assumed to reflect the sequential infection patterns in a community (\cite{fleming2015modelling}). Indeed, available data suggest that, when influenza and RSV are present in community outbreaks, they tend to affect all age groups simultaneously (\cite{dowell1996respiratory,hashem2003respiratory,MunywokiEtAl2013,wu2016coherence,zambon2001contribution}). Influenza surveillance data, such as from the \cite{AustraliaSurveillanceReport}, shows similar infection trajectories across different age groups.

\section{Modeling}\label{sec:Modeling}

This section describes a SKM of influenza and RSV dynamics as well as its diffusion approximation in the large-population limit, required to compute the likelihood of reported infection data and virological tests. We then construct a Bayesian model that relates the governing equations, including a background infection component, to the aggregated ARI reports and the virological test data.

\subsection{Stochastic kinetic model of a two-pathogen system}
\label{subsec:mathmodel}

Stochasticity appears in biological systems due to their discrete nature and the occurrence of environmental and demographic events. In the case of disease dynamics, the occurrence of events such as interactions between individuals that constitute exposure can be reasonably described as stochastic. Therefore, it is reasonable to model this stochasticity directly in the individual transitions, in contrast to indirectly modeling their aggregate behavior or perturbing a deterministic compartmental model. Stochastic kinetic or chemical master equation modeling (\cite{allen2008introduction,wilkinson2011stochastic}) is a mathematical formulation of Markovian stochastic processes which describe the evolution of the probability distribution of finding the system in a given state at a specified time (\cite{gillespie2007stochastic,thomas2012intrinsic}).

\begin{figure}
\centering
\begin{tikzpicture}[node distance=2.5cm,auto,>=latex']
\node [int,fill=white] (SS) {$S$};
\node [int] (SI) [right of=SS] {$X_{\mathrm{SI}}$};
\node [int] (SR) [right of=SI] {$X_{\mathrm{SR}}$};
\node [int] (IS) [below of=SS] {$X_{\mathrm{IS}}$};
\node [int] (RS) [below of=IS] {$X_{\mathrm{RS}}$};
\node [int] (IR) [below of=SR] {$X_{\mathrm{IR}}$};
\node [int] (RR) [below of=IR] {$X_{\mathrm{RR}}$};
\node [int] (RI) [right of=RS] {$X_{\mathrm{RI}}$};
\node [int, pin={[init]above left:$\mu $}](SS) {$X_{\mathrm{SS}}$};
\node [coordinate] (end) [right of=SR, node distance=2cm]{};

\draw [->] (SS) -- (1.0,-1.0) [above right] node {$\mu $};
\draw [->] (SI) -- (3.5,-1.0) [above right] node {$\mu $};
\draw [->] (SR) -- (6.0,-1.0) [above right] node {$\mu $};
\draw [->] (IS) -- (1.0,-3.5) [above right] node {$\mu $};
\draw [->] (RS) -- (1.0,-6.0) [above right] node {$\mu $};
\draw [->] (RI) -- (3.5,-6.0) [above right] node {$\mu $};
\draw [->] (RR) -- (6.0,-6.0) [above right] node {$\mu $};
\draw [->] (IR) -- (6.0,-3.5) [above right] node {$\mu $};
\draw [->] (0.5,1.0) --(4.0,1.0) node [above,midway] {RSV};
\draw [->] (-1.3,-0.5) -- (-1.3,-5.0) node[above, midway,rotate=90] {Influenza\vphantom{$A_{A_{A_{A_{A_{A_{A_A}}}}}}$}};

\path[->] (SS) edge node {$\beta _{2} \lambda _{2}(t)$} (SI);
\path[->] (SI) edge node {$\gamma $} (SR);
\path[->] (SR) edge node {$\sigma _{1} \beta _{1} \lambda _{1}(t)$} (IR);
\path[->] (SS) edge [left] node {$\beta _{1} \lambda _{1}(t)$} (IS);
\path[->] (IS) edge [left] node {$\gamma $} (RS);
\path[->] (RS) edge node {$\sigma _{2} \beta _{2} \lambda _{2}(t)$} (RI);
\path[->] (RI) edge node {$\gamma $} (RR);
\path[->] (IR) edge node {$\gamma $} (RR);
\end{tikzpicture}
\caption{SIR model with two pathogens. $X_{kl}$ represents the number of
individuals in immunological status $k$ for pathogen 1 and status
$l$ for pathogen 2. Labels above the arrows represent the reaction rates
for each reaction type.}
\label{fig:model1}
\end{figure}

To model the relationship between influenza and RSV (henceforward called pathogens 1 and 2, respectively) during a single year, we consider a closed population of size $\Omega $, assumed to be well mixed and homogeneously distributed, where the individuals interacting in a fixed region can make any of $\mathcal{R}$ possible transitions. The stochastic ``Susceptible-Infected-Recovered'' (SIR) model with two pathogens (\cite{adams2007influence,kamo2002effect,Vasco2007}) is described by eight compartments, corresponding to distinct immunological statuses. Denote by $X_{kl}(t)$ the \textit{number of individuals} at time $t$ in immunological status $k\in \{S , I, R\}$ for pathogen 1 and immunological status $l\in \{S, I, R\}$ for pathogen 2. We omit the state $ X_{\mathrm{II}}$ from the model because, although simultaneous infection by both viruses is biologically possible, clinical studies have found that the frequency of coinfection with RSV and influenza is usually low (\cite{MeskillEtAl2017}). A graphical representation of this model is provided in Figure~\ref{fig:model1}.

\begin{table}
\caption{Parameters defining the two-pathogen SKM for influenza and RSV}
\begin{tabular}{llll}
\hline
{Parameter} & \multicolumn{1}{c}{Description} & {Value} \\
\hline
$\Omega $         & Average yearly population size               & $2.5\times 10^{6}$              \\
$\mu $            & Birth and death rate                         & $1/70$~$\mathrm{years} ^{-1}$   \\
$\gamma $         & Recovery rate                                & $365/7$~$\mathrm{years} ^{-1}$  \\
$\lambda _{1}(t)$ & Proportion of influenza-infected individuals & $(x_{is}(t)+x_{ir}(t))/\Omega $ \\
$\lambda _{2}(t)$ & Proportion of RSV-infected individuals       & $(x_{si}(t)+x_{ri}(t))/\Omega $ \\
$\beta _{1}$      & Baseline transmission rate for influenza     & --                              \\
$\beta _{2}$      & Baseline transmission rate for RSV           & --                              \\
$\sigma _{1}$     & Cross-immunity or enhancement for influenza  & --                              \\
$\sigma _{2}$     & Cross-immunity or enhancement for RSV        & --                              \\
\hline
\end{tabular}
\label{tab:parameters_mod}
\end{table}

A list of reaction rate parameters is provided in Table~\ref{tab:parameters_mod}. The constants $\beta _{1}$ and $\beta _{2}$ represent the contact transmission rate, which describes the flow of individuals from the susceptible group to a group infected with pathogen 1 and 2, respectively. In the context of ARI, the average recovery time is known to be relatively stable and lasts for approximately seven days (\cite{CDC}). Therefore, the rate, $\gamma $, at which infected individuals recover (move from infected to temporary immunity in the recovered category) is $1/7$ $\mathrm{days} ^{-1}$. Since the population is relatively stable over the years under study, we set the birth rate equal to the death rate $\mu $ in our transition model. We also assume an average life expectancy of $1/\mu =70$ years (\cite{WHO2017}). Variables $\lambda _{1}(t)$ and $\lambda _{2}(t)$ represent the proportion of the population infected with pathogens 1 and 2, respectively, at a given time $t$. Finally, to describe the interaction between influenza and RSV, we use the cross-immunity or cross-enhancement parameters $\sigma _{1}$ and $\sigma _{2}$. Cross-immunity for influenza is present when $0<\sigma _{1} <1$, indicating that the presence of RSV inhibits infection with influenza. A value of $\sigma _{1}=0$ confers complete protection against influenza while a value of $\sigma _{1}=1$ confers no protection, and a value of $\sigma _{1}>1$ represents increasing degree of cross-enhancement, indicating that the presence of RSV enhances infection with influenza (\cite{adams2007influence}). The definition of $\sigma _{2}$ is analogous. Further customization of the transition mechanism is possible within this stochastic dynamical model (e.g., \cite{wilkinson2006stochastic}) and is desirable when additional information about the system dynamics is available. Examples include considering the effect of vaccination programs, seasonal forcing, or the presence of specific additional pathogens in circulation. We discuss these in the section on future work.

\begin{table}
\caption{Stoichiometric vectors and propensities for each transition type}
\label{tab:reactions}
\begin{tabular}{cccc}
\hline
\multicolumn{1}{l}{Index $j$} & Transition & Stoichiometric vector $v_{j}$ & Propensity $a_{j}(x)$ \\
\hline
1 & $0 \rightarrow X_{\mathrm{SS}}$ & $(1,0,0,0,0,0,0,0)^{\top }$ & $\mu \Omega $ \\
2 & $X_{\mathrm{SS}}\rightarrow X_{\mathrm{SI}}$ & $(-1,0,0,1,0,0,0,0)^{\top }$ & $\beta _{2}\lambda _{2} x_{ss}$ \\
3 & $X_{\mathrm{SS}}\rightarrow X_{\mathrm{IS}}$ & $(-1,1,0,0,0,0,0,0)^{\top }$ & $\beta _{1}\lambda _{1} x_{ss} $\\
4 & $X_{\mathrm{SS}}\rightarrow 0$ & $(-1,0,0,0,0,0,0,0)^{\top }$ & $\mu x_{ss}$ \\
5 & $X_{\mathrm{IS}}\rightarrow 0$ & $(0,-1,0,0,0,0,0,0)^{\top }$ & $\mu x_{is}$\\
6 & $X_{\mathrm{IS}}\rightarrow X_{\mathrm{RS}}$ & $(0,-1,1,0,0,0,0,0)^{\top }$ & $\gamma x_{is}$ \\
7 & $X_{\mathrm{RS}}\rightarrow 0$ & $(0,0,-1,0,0,0,0,0)^{\top }$ & $\mu x_{rs}$ \\
8 & $X_{\mathrm{RS}}\rightarrow X_{\mathrm{RI}}$ & $(0,0,-1,0,1,0,0,0)^{\top }$ & $\sigma \beta _{2}\lambda _{2}x_{rs} $\\
9 & $X_{\mathrm{SI}}\rightarrow X_{\mathrm{SR}}$ & $(0,0,0,-1,0,1,0,0)^{\top }$ & $\gamma x_{si}$ \\
10 & $X_{\mathrm{SI}}\rightarrow 0$ & $(0,0,0,-1,0,0,0,0)^{\top }$ & $\mu x_{si}$ \\
11 & $X_{\mathrm{RI}}\rightarrow X_{\mathrm{RR}}$ & $(0,0,0,0,-1,0,0,1)^{\top }$ & $\gamma x_{ri}$ \\
12 & $X_{\mathrm{RI}}\rightarrow 0$ & $(0,0,0,0,-1,0,0,0)^{\top }$ & $\mu x_{ri}$ \\
13 & $X_{\mathrm{SR}}\rightarrow 0$ & $(0,0,0,0,0,-1,0,0)^{\top }$ & $\mu x_{sr}$ \\
14 & $X_{\mathrm{SR}}\rightarrow X_{\mathrm{IR}}$ & $(0,0,0,0,0,-1,1,0)^{\top }$ & $\sigma \beta _{1}\lambda _{1}x_{sr}$\\
15 & $X_{\mathrm{IR}}\rightarrow 0$ & $(0,0,0,0,0,0,-1,0)^{\top }$ & $\mu x_{ir}$ \\
16 & $X_{\mathrm{IR}}\rightarrow X_{\mathrm{RR}}$ & $(0,0,0,0,0,0,-1,1)^{\top }$& $\gamma x_{ir}$ \\
17 & $X_{\mathrm{RR}}\rightarrow 0$ & $(0,0,0,0,0,0,0,-1)^{\top }$ & $\mu x_{rr}$ \\
\hline
\end{tabular}
\end{table}

We next make the following standard assumptions about the random vector of infection states $X(t)=  (X_{\mathrm{SS}}(t), X_{\mathrm{IS}}(t), X_{\mathrm{SR}}(t), X_{\mathrm{RS}}(t), X_{\mathrm{SI}}(t), X_{\mathrm{RR}}(t), X_{\mathrm{RI}}(t), X_{\mathrm{IR}}(t)  )^{\top }$ at time $t$ (see, e.g., \cite{allen2008introduction,gillespie2007stochastic}). For an infinitesimally small time increment, $\Delta t$, the probability of the occurrence of an event of type $j$ in the interval $(t, t+\Delta t]$ is $a_{j}(x)\Delta t$, where $a_{j}(x)$ denotes the propensity function for the reaction at state $X = x$. Under these conditions, and assuming a well-mixed population, the probability mass function $p_{t}$ describing the probability of being in state $X= x$ at time $t$ evolves according to the Kolmogorov forward equation (chemical master equation, or CME),
%
\begin{equation}
\label{eqs:ME} \frac{dp_{t}(x)}{dt} = \sum_{j=1}^{\mathcal{R}}
\bigl\{ a_{j}(x-v_{j})p_{t}(x-v_{j})-
a_{j}(x)p_{t}(x) \bigr\} ,
\end{equation}
where $v_{j}(t)$ are stoichiometric vectors whose elements in $\{-1, 0 ,1\}$ describe the addition or subtraction of individuals from a particular compartment of $x$. Values of $a_{j}(x)$ and $v_{j}(t)$ for the two-pathogen SIR model are provided in Table~\ref{tab:reactions}. The large-population approximation to this system characterizes the distribution of the Markov process $X(t), t\in [0,T]$ as the sum of a deterministic term $\Omega \phi (t),   \phi : [0, T]\to \mathbb{R}^{+\dim (X)}$ and a stochastic term $\Omega ^{1/2}\xi $, where $\xi $ is a Gaussian process with mean $\tilde{\xi }(t): [0, T]\to \mathbb{R}^{+\dim (X)}$ and covariance $C(t): [0, T]\to \mathbb{R}^{\dim (X) \times \dim (X)}$. Equivalently, we can write,
%
\begin{align}
X(t) \sim {\mathcal{N}} \bigl(\Omega \phi (t)+\Omega ^{1/2}\tilde{\xi
}(t), \Omega C(t) \bigr). \label{lik:process}
\end{align}
The \hyperref[Appendix]{Appendix} explains this large-population approximation and defines the quantities $\phi $, $\tilde{\xi }$, $C$ as solutions to ODE initial value problems involving the propensities $a_{j}$ and stoichiometric matrix $V$ for the system. Section~\ref{subsec:posterior} describes how this approximation is used to model aggregated data on ARI reports.

\subsection{Probability models for aggregate reports and virological tests}
\label{subsec:posterior}

Our first data set consists of indirect observations of the Markov process $X(t): t\in [0,T]$ transformed via $G^{T} = [0,1,0,1,1,0,1,0]$. That is, observations are made on the total number $G^{\top }X(t) = X_{\mathrm{IS}}(t)+X_{\mathrm{IR}}(t)+X_{\mathrm{SI}}(t)+X_{\mathrm{RI}}(t)$ of reported infections from influenza and RSV at $M$ discrete observation time points $t_{1}, \ldots , t_{M}$. The following distributional fact follows from (\ref{lik:process}) and the standard assumption (e.g., \cite{fearnhead2014inference}) of a zero initial drift term, $\tilde{\xi }(0)=0$, so that $\tilde{\xi }(t)=0$ for all $t\in [0, T]$. Thus,
%
\begin{equation}
G^{\top }X(t_{i}) \mid \theta \sim {\mathcal{N}} \bigl(\Omega
G^{\top }\phi (t_{i}), \Omega G^{\top }C(t_{i})G
\bigr), \label{eqs:marginal_likelihood}
\end{equation}
where $\theta $ is a vector of the SKM parameters, defined in Section~\ref{subsec:mathmodel}, augmented with unknown initial conditions $X(0)$,
%
\begin{align}
\theta = \bigl(\beta _{1},\beta _{2},\sigma
_{1}, \sigma _{2}, X(0)\bigr).
\end{align}

The aggregate number of ARI cases in the San Luis Potos\'{i} data also includes infections by viruses other than influenza and RSV. Although these may be responsible for a significant fraction of all ARI cases, influenza and RSV are the two viruses that drive the epidemic fluctuations observed during the winter outbreaks in each year. Other viruses in circulation are included in the model by augmenting the state vector $X$ by a background process corresponding to ARI infection that is not caused by influenza or RSV. In contrast to infection states for influenza and RSV, this background state $D$, which represents the number of individuals infected with a different pathogen, is modeled empirically. We assume that the number of background infections in a given week and the previous week are positively correlated and that the variance of the process error for $D$ lies between the variance of the infection state $X$ (proportional to $\Omega $, as in (\ref{lik:process})) and the variance of the observation error (proportional to $\Omega ^{2}$). This choice ensures that the background process is more variable than $X$ but is not flexible enough to overfit the data. A positive drift term, proportional to $\Omega $, reflects the expected growth in background cases. This suggests the lagged background model,
%
\begin{align}
D_{i} = \Omega c + \nu D_{i-1} + \eta _{i},
\label{eqn:ar1}
\end{align}
where $\eta _{i} \sim \mathcal{N}  (0,\Omega ^{3/2}\kappa   )$ are independent, zero-mean, stochastic noise terms, $c$, $\nu $, and $\kappa $ are unknown positive constants, and $D_{1} \sim \mathcal{N}  (\Omega c,\Omega ^{3/2}\kappa   )$ is the distribution of the initial background state. Discrepancy models, such as this one, must both be flexible enough to capture the unmodeled signal but structured enough to guard against overfitting (e.g., \cite{Brynjarsd2014}). In our model this is accomplished by setting the process error variance $\Omega ^{3/2}\kappa $ to be below that of the observation error but above that of $X$ (our choice of $\kappa = c_{0}$ makes these variances directly comparable). Furthermore, the integration of the additional virological test data in the analysis helps to constrain the background model.

Observed aggregate counts are assumed subject to independent Normal observation errors $\varepsilon _{i}$ with mean zero and variance $\Omega ^{2}\Sigma $. We make the additional assumption that a fixed proportion, $r\in [0,1]$, of all individuals infected with ARI seek consultation. Therefore, the observed aggregated reports at observation times $t_{1},\ldots , t_{M}$ are modeled as
%
\begin{align}
Y_{i} = r \bigl\{ G^{\top }X(t_{i}) +
D_{i} \bigr\} + \varepsilon _{i}. \label{lik:Y}
\end{align}
The justification for modeling the total number of influenza and RSV infections directly via the Normal model, obtained via the LNA, is computational (\cite{fearnhead2014inference,Golightly2015}) by exploiting the availability of closed-form Bayesian updates over the states. Modeling counts, via a discrete distribution centered at the LNA, would require an additional layer of sampling at each observation location for each Markov chain Monte Carlo (MCMC) iteration which would quickly become computationally infeasible.

Due to symmetry in the two-pathogen SIR model (Figure~\ref{fig:model1}), conditioning on aggregate counts (\ref{lik:Y}) alone leaves uncertainty about the circulation patterns of individual pathogen infection trajectories. Our analysis shows that the inclusion of the virological data, described in Section~\ref{sec:data}, can aid in disaggregating the dynamics of the two pathogens. Denote by $N_{1,i}$, $N_{2,i}$, $N_{3,i}$ the number of virological laboratory tests that identified influenza, RSV, and neither pathogen, respectively, for children under five years of age admitted with an ARI during week $i$ at Hospital Central. The total number of virological tests is related to the number of infected individuals in the population as follows. According to the 2010 census, there were $\Omega _{c} = 266\text{,}761$ children under five years of age in San Luis Potos\'{i} out of a total population of $\Omega = 2\text{,}585\text{,}518$ (Consejo Nacional de Poblaci\'{o}n, M\'{e}xico). Furthermore, available data suggests that approximately a proportion $r_{h} = 0.0005$ of children with reported ARI in the state are admitted to Hospital Central and are, therefore, eligible for a virological test. Thus, we model the expected number of virological tests that were positive for influenza, RSV, and other background infections, respectively, as
\begin{align*}
m_{1,i} &= \mathbf{E} \bigl[N_{1,i}\mid X(t_{i}),
\theta , \tau \bigr] = r_{h} r (\Omega _{c}/\Omega ) \bigl\{
X_{\mathrm{IS}}(t_{i})+X_{\mathrm{IR}}(t_{i}) \bigr\} ,
\\
m_{2,i} &= \mathbf{E} \bigl[N_{2,i}\mid X(t_{i}),
\theta , \tau \bigr] = r_{h} r (\Omega _{c}/\Omega ) \bigl\{
X_{\mathrm{SI}}(t_{i})+X_{\mathrm{RI}}(t_{i}) \bigr\} ,
\\
m_{3,i} &= \mathbf{E} \bigl[N_{3,i}\mid X(t_{i}),
D_{i}, \theta , \tau \bigr] = r_{h} r (\Omega
_{c}/\Omega ) D_{i}.
\end{align*}
The variability in the true proportion of the infected population admitted to Hospital Central, motivates the use of the variance-inflated model,
%
\begin{align}
N_{k,i}\mid X(t_{i}), D_{i}, \theta ,
\tau \sim \operatorname{NegBin} \bigl(v m_{k,i},1/\bigl(1+v^{-1}
\bigr) \bigr), \quad k = 1, 2, 3, \label{eqn:infectionprop}
\end{align}
where $v$ is an unknown variance inflation parameter. The variance of (\ref{eqn:infectionprop}) is $m_{k,i}  (1 + 1/v  )$ and is controlled by the parameter $v$ which will be estimated along with the other model components.

\begin{table}
\centering
\caption{Parameters defining the observation and background models}\label{tab:parameters_err}
\begin{tabular}{ll}
\hline
{Parameter} &\multicolumn{1}{c}{Description} \\
\hline
$c$ & Constant part of drift term for the number of background infections \\
$\nu $ & Lag coefficient for the number of background infections \\
$\Sigma $ & Constant part of error variance for reported infections\\
$r$ & Reporting proportion for those infected with an ARI\\
$r_{h}$ & Average proportion of children with ARI admitted to the Hospital Central\\
$\Omega _{c}$ & Average yearly number of children under 5 years of age in San Luis Potos\'{i} \\
$v$ & Variance inflation factor for virological data \\
\hline
\end{tabular}
\normalsize
\end{table}

Parameters defining the discrepancy model (\ref{eqn:ar1}) and observation processes (\ref{lik:Y}) and (\ref{eqn:infectionprop}) are listed in Table~\ref{tab:parameters_err}. In the following we denote by $\tau $ the subset of these parameters which are unknown:
\begin{align*}
\tau = (c, \nu , r, v, \Sigma ).
\end{align*}
Furthermore, we assume that both the SKM parameters $\theta $ and the discrepancy and observation model parameters $\tau $ vary across epidemic years. This is due to factors, such as the circulation of different strains of ARI in different years, which are difficult to accurately model. The rationale for this assumption is further explained in the Discussion.

\subsection{Prior probability model for unknown parameters}
\label{sec:prior}

Prior distributions on the model and auxiliary parameters are obtained by expert elicitation and by enforcing physical constraints. For example, ensuring that initial conditions on the states represent the entire population requires that elements of the unknown initial state vector $X(0)/\Omega $ lie on the simplex, motivating the use of a Dirichlet prior. The reporting proportion $r\in (0,1)$ and the small positive lag coefficient $\nu $ for the background process are assigned uniform prior distributions on $(0,1)$. The remaining parameters are bounded below by zero and are, therefore, assigned Gamma priors. The shape and scale hyperparameters for the Gamma prior on the transmission rates $\beta _{p}, p=1,2$ are chosen to be 20 and 3, respectively, to yield a reasonable prior mean of 60 $\mathrm{days} ^{-1}$ and a large spread, reflecting our uncertainty about these quantities. The shape and scale for the Gamma prior on the cross-interference parameters $\sigma _{p}, p =1,2$, are chosen to be 10 and 0.1, respectively. This choice places a relatively large prior weight on the region around the neutral case $\sigma _{p} = 1$, $p=1,2$. The same hyperparameters are chosen for the variance inflation factor $v$ so that the prior mean corresponds to a relatively small spread in the counts of virological test data. The prior on the error variance $\Sigma $ is Gamma with shape parameter $1$ and scale parameter $0.01$, ensuring a moderately sized prior mean on the scale of the population and substantial posterior mass near zero. The state prior covariance, $C_{0} = c_{0} \mathbb{I}_{\dim (X),\dim (X)}$, is assumed to be diagonal with prior variance $c_{0}$ and assigned a moderate value of 0.01.

\section{Inferential methodology}\label{sec:methodology}

Consider inference on a general partially-observed SKM with states $X$ and model parameters $\theta $. A Markov chain Monte Carlo (MCMC) algorithm, targeting the posterior $\pi (\theta \mid y)$, requires the expensive evaluation of the marginal likelihood $p  (y \mid \theta   )$ at every iteration given $\theta $. An approximate but more efficient evaluation of the marginal likelihood can be obtained via a linear filter under the LNA. The inference problem can then be formulated as a hidden Markov model (\cite{fearnhead2014inference,Finkenstdt2013,Golightly2015}). In this scenario the system equation for the hidden state (\ref{lik:process}) is
%
\begin{align}
 X(t_{i}) = FX(t_{i-1}) + H(t_{i}) +
\zeta (t_{i}), \quad \zeta (t_{i}) \sim N\bigl(0, \Omega
C(t_{i})\bigr), \label{eqn:hmm-system}
\end{align}
where $F$ is the zero $\dim (X)\times \dim (X)$ matrix and $H(t_{i})=\Omega \phi (t_{i})$. Here, $\phi (t_{i})$ and $C(t_{i})$ are obtained by integrating equations (\ref{eqs:_detpartODE}) and (\ref{eqn:SDEterm}) from time $t_{i-1}$ to $t_{i}$ with the initial condition $\phi (t_{i})$ and $C(t_{i})$, respectively (to generalize this formulation for the case of a nonzero drift term $\tilde{\xi }$, see \cite{Finkenstdt2013}). A linear Gaussian observation process allows the use of the computationally efficient linear filter. However, in our applied problem the observation process (\ref{lik:Y}) and (\ref{eqn:infectionprop}) is more complex, incorporating an additional non-Gaussian data source and a background process. Section~\ref{subsec:extendedpma} outlines an extended sampling approach to incorporate additional data and model components into this framework.

\subsection{Extended marginal sampling approach based on LNA}
\label{subsec:extendedpma}

Algorithm~\ref{algo:extendedsmc} summarizes an extended sampling procedure to be performed at each iteration of an MCMC targeting $\pi _{a}  ( \theta ,\tau , x_{1:M}, d_{1:M} \mid y_{1:M}, n_{1:M}   )$ given sampled values of $(\theta , \tau )$. We will use the shorthand notation $x_{1:i}$ to denote the vector $x$ at times $t_{1}$ through $t_{i}$. Incorporating the background process (\ref{eqn:ar1}) within this framework is straightforwardly done by augmenting equation (\ref{eqn:hmm-system}) to include infection states with ARI other than influenza and RSV. Formulation of this problem as a hidden Markov model enables the use of a backward recursion to compute the smoothing distribution over the epidemic state (e.g., \cite{Touloupou2019}). Once this is available, a likelihood-free sampling step over the states allows us to compute the likelihood component for the virological test data $N$.

Assuming conditional independence between $Y$ and $N$, given the infection states, background, and parameters, the posterior density,
%
\begin{equation}
\begin{aligned} & \pi _{a} ( \theta ,\tau ,
x_{1:M}, d_{1:M} \mid y_{1:M}, n_{1:M} )
\\
&\quad \propto p (n_{1:M} \mid x_{1:M}, d_{1:M},
\theta , \tau ) p_{a} (x_{1:M}, d_{1:M} \mid
y_{1:M}, \theta , \tau ) p_{a} (y_{1:M} \mid \theta
, \tau ) \pi (\theta ,\tau ), \end{aligned} \label{eqn:post}
\end{equation}
is proportional to the prior multiplied by the likelihood components (\ref{lik:process}), (\ref{eqn:ar1}), (\ref{eqn:infectionprop}), and the approximate marginal likelihood $p_{a}  (y_{1:M} \mid \theta ,\tau   )$ obtained by alternating forecast and analysis steps (see, e.g., \cite{HarrisonWest1997}, and Algorithm~\ref{algo:extendedsmc}). The forecast and analysis steps used to obtain $p_{a}  (y_{1:M} \mid \theta ,\tau   )$ provide the necessary information to obtain the smoothing distribution over $x_{1:M}$ and $d_{1:M}$ via backward recursion. Conditioning on a sample from this distribution, we can then evaluate $p  (n_{1:M} \mid x_{1:M}, d_{1:M}, y_{1:M}, \theta , \tau   )$. This approach has the additional advantage of allowing visualization of the posterior samples over the infection states and background.

\begin{algorithm}
\caption{
Sequential sampler implemented at each iteration of a MCMC algorithm targeting $\pi _{a}  ( \theta ,\tau , x_{1:M}, d_{1:M} \mid y_{1:M}, n_{1:M}  )$, enabling computation of the acceptance probability
}
\label{algo:extendedsmc}
\DontPrintSemicolon
Define the augmented state vector $\tilde{Z} = [X^{\top},D]^{\top}$ and observation vector $\tilde{G}=[G^{\top},1]^{\top}$. Define the matrix
$\tilde{V}_{i} =
\bigl[\begin{smallmatrix}
V_{i} & 0 \\
0 & V_{i}^{d}
\end{smallmatrix}\bigr]
$ initialized with $V_{0}= \Omega c_{0} \mathbb{I}_{\dim (X),\dim (X)}$ and $V_{0}^{d} = \Omega ^{3/2}\kappa $. Define the vector $\tilde{m}_{i}= [m_{i},m^{d}_{i}]^{\top}$ initialized with $m_{0} = x^{\top}_{1}$ and $m^{d}_{0} = \Omega c$.  Compute prior marginal likelihood:
$$p_{a}(y_{1}\mid \theta , \tau ) = \mathcal{N}  (y_{1}; r \tilde{G}^{\top}\tilde{m}_{0}, r^{2} \tilde{G}^{\top} \tilde{V}_{0} \tilde{G} + \Omega ^{2}\Sigma   ).$$

The analysis distribution at time $t=1$ is $p  (\tilde{z}_{1}\mid y_{1}  ) = \mathcal{N}  (\tilde{z}_{1}; \tilde{m}_{1},\tilde{V}_{1}  )$, where
\begin{align*}
\tilde{m}_{1} &= \tilde{m}_{0} + r\tilde{G}^{\top}
\tilde{V}_{0}\bigl(r\tilde{G}^{T} \tilde{V}_{0}
\tilde{G} + \Omega ^{2}\Sigma \bigr)^{-1}
\bigl(y_{1}-r\tilde{G}^{T} \tilde{m}_{0} \bigr),
\\
\tilde{V}_{1} &= \tilde{V}_{0}- r^{2}
\tilde{G}^{\top}\tilde{V}_{0}\bigl(r^{2}
\tilde{G}^{\top}\tilde{V}_{0}\tilde{G}+\Omega ^{2}
\Sigma \bigr)^{-1}\tilde{V}_{0}\tilde{G}.
\end{align*}

\For{$i=1,\ldots ,M-1$}{

Retain the previously computed quantities $\tilde{m}_{i}$,  $\tilde{V}_{i}$, $p_{a}  (y_{1:i} \mid \theta , \tau   )$. Integrate equations (\ref{eqs:_detpartODE}) and (\ref{eqn:SDEterm}) from time $t_{i}$ to $t_{i+1}$ using as initial conditions $\phi _{i} = m_{i}/\Omega $ and $C_{i} = V_{i}/\Omega $, to obtain $\phi _{i+1}$ and $C_{i+1}$.  Augment $\tilde{\phi}_{i+1} = [\phi _{i+1}^{\top}, c + \nu m^{d}_{i}/\Omega ]^{\top}$ and $\tilde{C}_{i+1}=
\bigl[\begin{smallmatrix}
C_{i+1} & 0 \\
0 & \Omega ^{1/2}\kappa
\end{smallmatrix}\bigr]$. 
 The filtering distribution of $\tilde{Z}$ is
$$p_{a}  (\tilde{z}_{i+1}\mid y_{1:i}  ) = \mathcal{N}  (\tilde{z}_{i+1}; \Omega \tilde{\phi}_{i+1}, \Omega \tilde{C}_{i+1}  ).$$

Compute the marginal likelihood of $Y_{1:(i+1)}$,
$$
p_{a}(y_{1:(i+1)}\mid \theta , \tau )= p_{a}  (y_{1:i} \mid \theta , \tau   )   p_{a}  (y_{i+1} \mid y_{1:i}  ),
$$
where,
$p_{a}  (y_{i+1} \mid y_{1:i}  ) = \mathcal{N}  ( y_{i+1}; r\Omega \tilde{G}^{\top} \tilde{\phi}_{i+1}, r^{2}\Omega \tilde{G}^{\top}\tilde{C}_{i+1}\tilde{G} + \Omega ^{2}\Sigma   )$

The approximate analysis distribution at time $t_{i+1}$ is $p_{a}  (\tilde{z}_{i+1} \mid y_{1:i+1}  ) =  \mathcal{N}  (\tilde{z}_{i+1}; \tilde{m}_{i+1},\tilde{V}_{i+1}  )$, where
\begin{align*}
\tilde{m}_{i+1} &= \Omega \phi _{i+1} + r\Omega
\tilde{G}^{\top}\tilde{C}_{i+1}\bigl(r\Omega
\tilde{G}^{T} \tilde{C}_{i+1} \tilde{G} + \Omega
^{2}\Sigma \bigr)^{-1} \bigl(y_{i+1}-r\Omega
\tilde{G}^{T} \phi _{i+1} \bigr),
\\
\tilde{V}_{i+1} &= \Omega \tilde{C}_{i+1}- r^{2}
\Omega ^{2}\tilde{G}^{\top}\tilde{C}_{i+1}
\bigl(r^{2}\Omega \tilde{G}^{\top}\tilde{C}_{i+1}
\tilde{G}+\Omega ^{2}\Sigma \bigr)^{-1}\tilde{C}_{i+1}
\tilde{G}.
\end{align*}

Backward recursion, together with the approximation used to obtain the analysis distribution above, leads to a smoothing distribution for $\tilde{Z}_{1:M}$  equal to the approximate analysis distribution,
$$p_{a}  (\tilde{z}_{1:M} \mid y_{1:M}  ) =  \mathcal{N}  (\tilde{z}_{1:M};   \tilde{m}_{1:M},\tilde{V}_{1:M}  )$$
(see, e.g., \cite{HarrisonWest1997}).
Sample a realization of $\tilde{z}_{1:M} = [x_{1:M}, d_{1:M}]$ from $p_{a}  (\tilde{z}_{1:M} \mid y_{1:M}  )$, and compute the likelihood component $p  (n_{1:M} \mid x_{1:M}, d_{1:M}, y_{1:M}, \theta , \tau   )$ from (\ref{eqn:infectionprop}).

Return the product $p_{a}  (y_{1:M} \mid \theta , \tau   )p  (n_{1:M} \mid x_{1:M}, d_{1:M}, y_{1:M}, \theta , \tau   )$, and store the associated sample paths $x_{1:M}$ and $d_{1:M}$ for posterior visualization of the infection trajectories.}
\end{algorithm}

\section{Results}\label{sec:results}

This section describes the results of the analysis of data from San Luis Potos\'{i}. Retrospective joint inference for the dynamics of influenza and RSV is carried out independently for each epidemic year due to the presence of yearly variability between circulating strains of ARI infections, as explained in the Discussion.

\subsection{Computational implementation}
\label{subsec:comp}

Posterior sampling was conducted by embedding the extended marginal sampling algorithm \ref{algo:extendedsmc} within a parallel tempering Markov chain Monte Carlo sampler (PTMCMC, \cite{Geyer1991}) over $(\theta , \tau , x_{1:M}, d_{1:M})$. PTMCMC is well suited to the present setting, because it can effectively explore challenging posterior densities, with features such as ripples and ridges which occur when different combinations of parameters lead to similar model fits. Five hundred thousand iterations of PTMCMC were obtained for each year using four parallel chains each. Due to strong posterior correlations between groups of parameters in our model, the parameters were sampled in three distinct blocks. The first two blocks consisted of the model parameters $\beta _{1}$, $\beta _{2}$, $\sigma _{1}$, $\sigma _{2}$, and parameters defining the discrepancy and observation processes, $\Sigma $, $r$, $c$, $v$, $\nu $. Proposal covariance adaptation was performed for both blocks, during the first half of each chain, which was discarded as burn-in. The last block consisted of the initial states $X_{\mathrm{SS}}(0)$, $X_{\mathrm{IS}}(0)$, $X_{\mathrm{RS}}(0)$, $X_{\mathrm{SI}}(0)$, $X_{\mathrm{RI}}(0)$, $X_{\mathrm{SR}}(0)$, $X_{\mathrm{IR}}(0)$, $X_{\mathrm{RR}}(0)$. For this block, only the proposal variances were adapted, because both the sign and magnitude of posterior correlations between these quantities vary across the parameter space. Convergence was diagnosed by visual inspection of trace plots and correlation plots.

\subsection{Inference based on data from San Luis Potos\'i, M\'exico}

\begin{figure}
\centering
\includegraphics[width=1\textwidth]{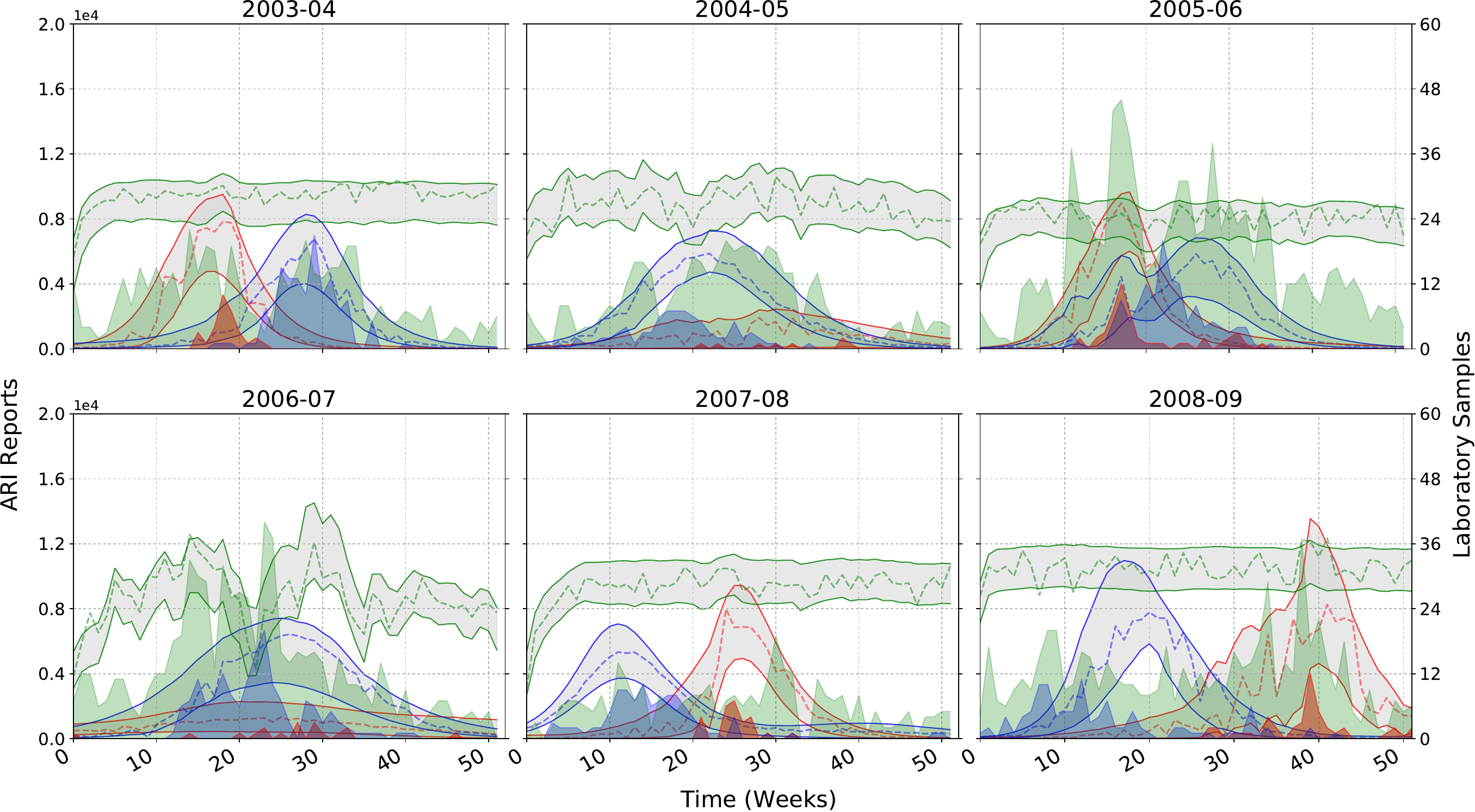}
\caption{Posterior summaries for disaggregated reports based on data from San Luis Potos\'{i}, M\'{e}xico. Left axis: Grey bands represent 95\% credible intervals for influenza (red outline), RSV (blue outline), and background infections (green outline). Dashed lines represent maximum a-posteriori estimated trajectories. Right axis: Reports of influenza and RSV from ARI affected children under five years of age at Hospital Central with tests positive for influenza, RSV, and neither are shown as shaded red, blue, and green polygons, respectively. Corresponding posterior predictive summaries are shown in Figure~1 of the Supplementary Material.}
\label{fig:real-data-posterior-sample-paths-disaggregated-all-years}
\end{figure}

\begin{figure}
\centering
\includegraphics[width=1\textwidth]{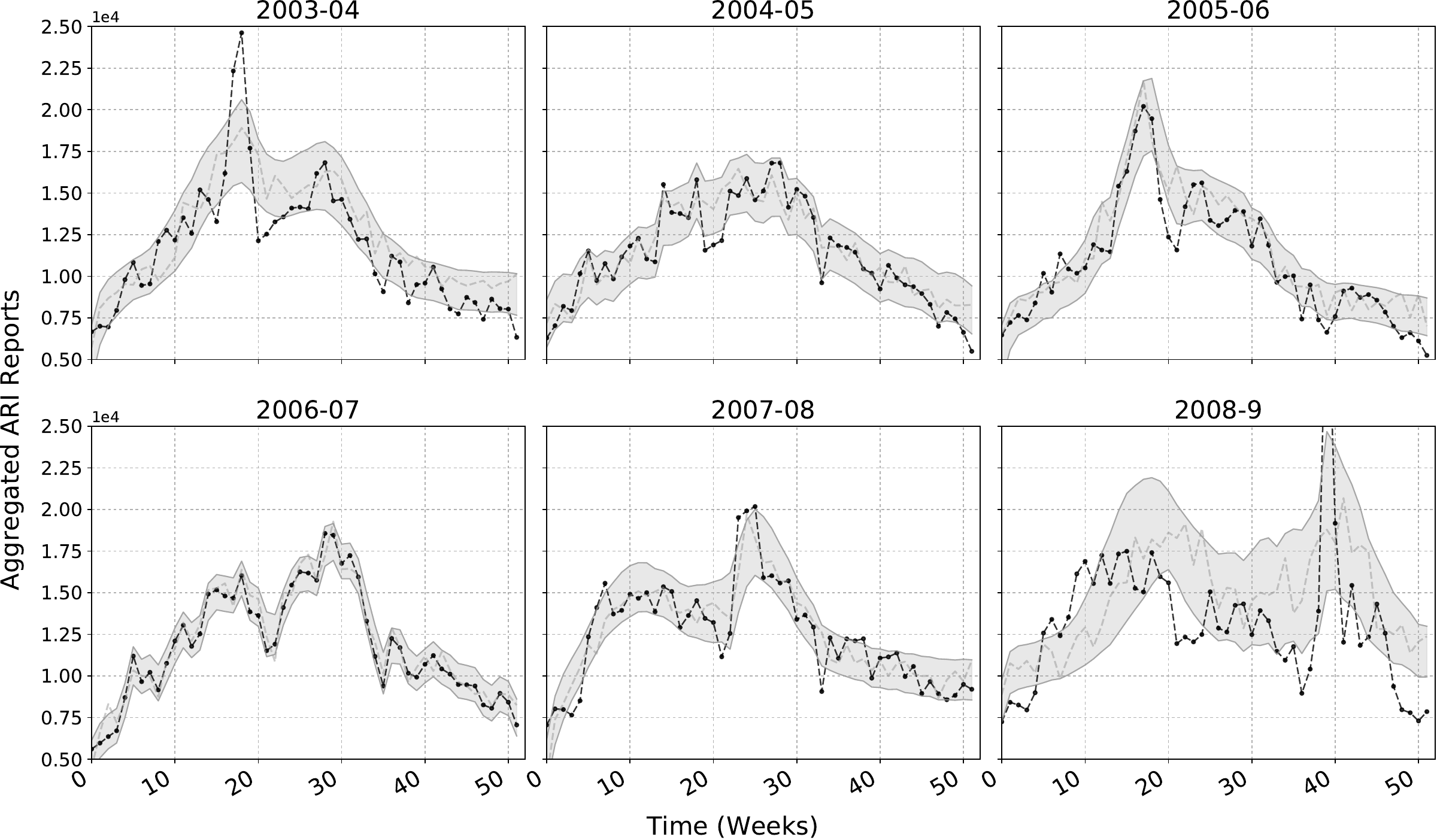}
\caption{Posterior summaries for aggregated ARI reports from San Luis Potos\'{i}, M\'{e}xico. The aggregated data, measured from August to July of the following year, is represented by black dots. Gray bands represent pointwise 95\% credible intervals, and the gray dashed line shows a maximum a-posteriori estimated trajectory. Corresponding posterior predictive summaries are shown in Figure~2 of the Supplementary Material.}
\label{fig:real-data-posterior-sample-paths-all-years}
\end{figure}

This section presents and interprets the results for retrospective joint inference for the dynamics of influenza and RSV from observed aggregate ARI counts and auxiliary virological testing data from San Luis Potos\'{i}, M\'{e}xico. Maximum a-posteriori (MAP) estimates and 95\% credible intervals for the model parameters are provided in Tables~1 through 6 of the Supplementary Material (\cite{ChkrebtiiEAlSupp2021}). Summaries of the marginal posterior distribution over ARI reports are shown in Figures~\ref{fig:real-data-posterior-sample-paths-disaggregated-all-years} and \ref{fig:real-data-posterior-sample-paths-all-years}.

Figure~\ref{fig:real-data-posterior-sample-paths-disaggregated-all-years} illustrates marginal posterior summaries over disaggregated infection states (influenza, RSV, and background). In all the years examined, except 2006--07, a single pathogen, either influenza or RSV, peaks first, followed by a peak associated with the other pathogen. Regardless of which peaks first, in all six years the posterior trajectories agree with the patterns suggested by the virological test data. Specifically, in all years, except 2003--04 and 2005--06, RSV cases peaks first, followed by influenza, with infections subsiding by the end of the epidemic year. In 2005--06, it appears that a small initial outbreak of RSV coincides with the main outbreak of influenza which is also consistent with the virological test data for that year. Secondary peaks that sometimes appear in the number of infections by a given pathogen may be due to local outbreaks of the ARI or ones caused by a different strain of the pathogen. The flexibility of our model to capture multiple epidemic peaks is useful and will be further considered in light of recent literature in the Discussion. The background model, which describes discrepancy due to other ARI infections as well as some degree of model misspecification, appears to suggest possible over-estimation of the number of infected individuals at the beginning of each epidemic season, due to the inability of the SKM to reproduce the observed dynamics starting from a small number of initial infections. This effect may be due to exogenous factors, including the fact that the beginning of the epidemic season (August) coincides with the beginning of the school year, when students come into more frequent contact with one another.

Figure~\ref{fig:real-data-posterior-sample-paths-all-years} aggregates these summaries to allow for visual comparison with the data. Posterior predictive plots are provided in the Supplementary Material as an additional tool to assess coverage. Figure~\ref{fig:real-data-posterior-sample-paths-all-years} suggests that the model is not flexible enough to account for very steep local peaks in infection. These peaks could be due to exogenous factors that were not accounted for by the model, such as school schedules, weather, or more than one strain of influenza and RSV in circulation. In particular, the year 2008--09 was characterized by the outbreak of a second, highly virulent strain of influenza A (H1N1) which may be responsible for the poorer model fit. In contrast, during the years 2004--05 and 2006--07, the combination of the ARI dynamics and background process track the data quite well. Indeed, both of these years exhibit slowly-changing ARI dynamics, which are consistent with our process model, and the lack of sharp peaks means that the background process is able to capture most of the remaining variability.

\begin{figure}
\centering
\includegraphics[width=0.475\textwidth]{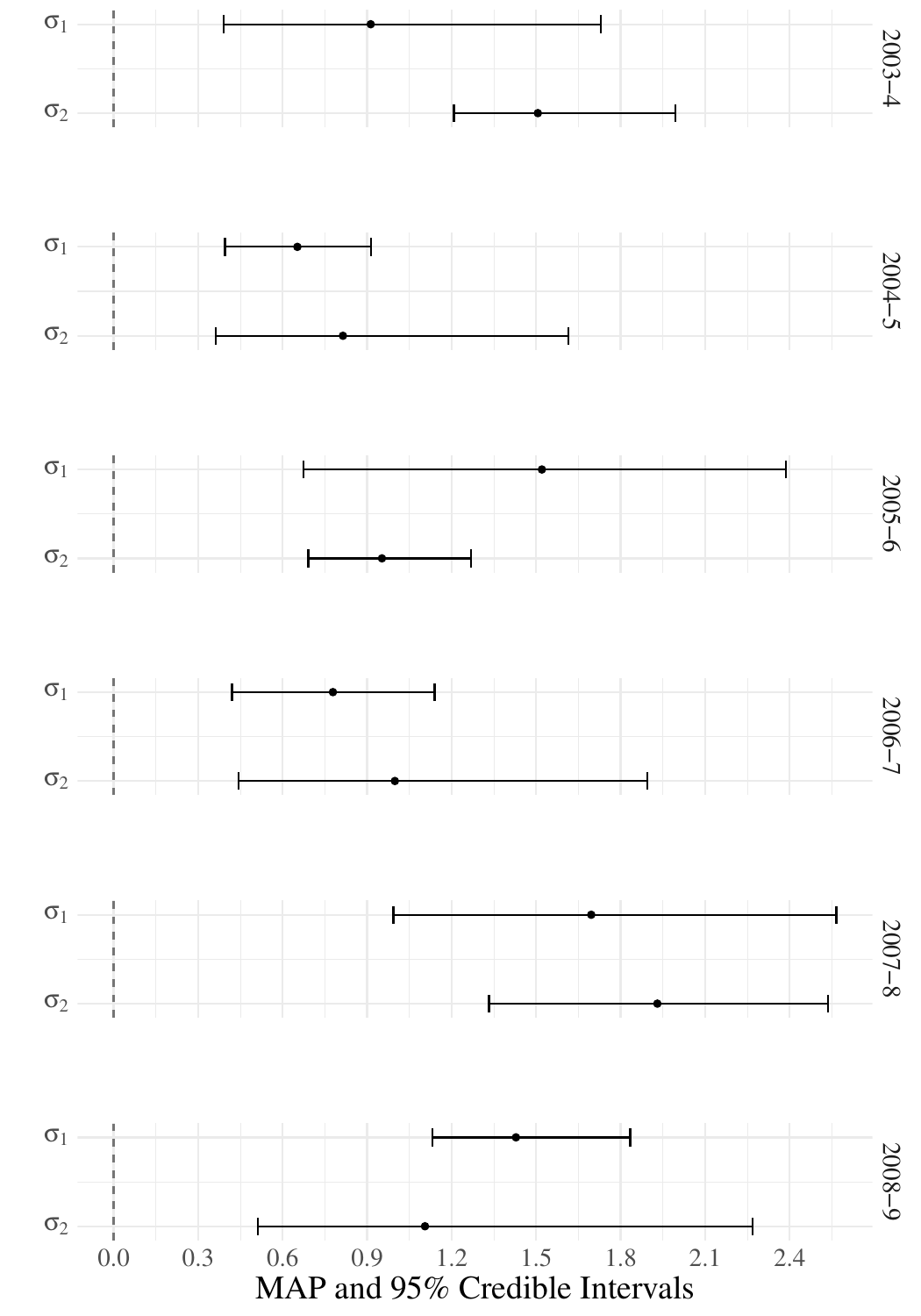}
\includegraphics[width=0.475\textwidth]{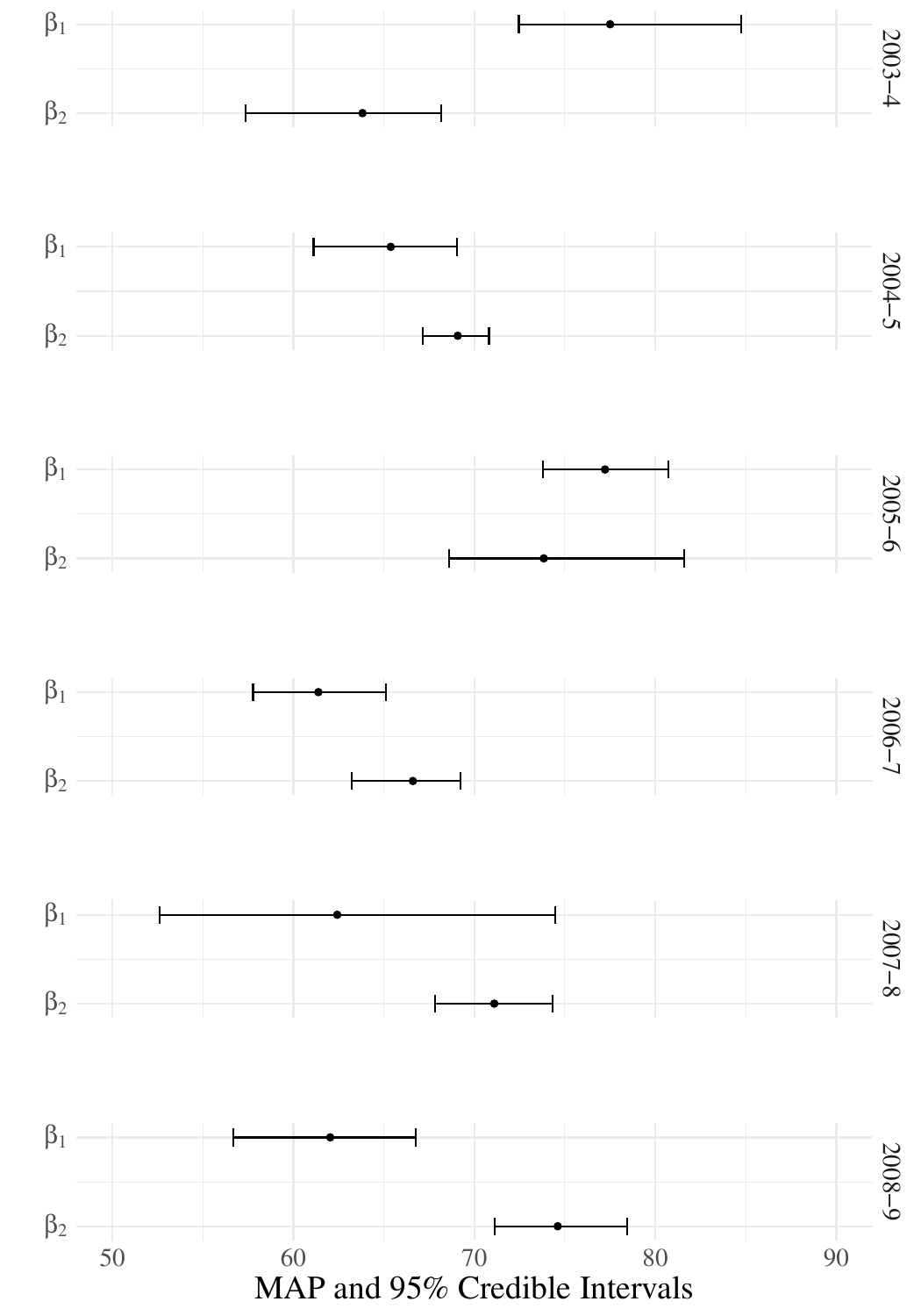}
\caption{Posterior summaries for each epidemic year (rows) of the cross-interference parameters $\sigma _{1}$ and $\sigma _{2}$ (left panel) and contact transmission rates $\beta _{1}$ and $\beta _{2}$ (right panel) for influenza and RSV, respectively, in San Luis Potos\'{i}, M\'{e}xico. Dots denote MAP estimates of the parameters for each year, and the bars represent the 95\% credible intervals.}
\label{fig:real-data-posterior-samples-beta-sigma}
\end{figure}

Estimates and credible intervals for cross-interference parameters $\sigma _{1}$ and $\sigma _{2}$ are shown in the left panel of Figure~\ref{fig:real-data-posterior-samples-beta-sigma}. Our results suggest that the degree of cross-immunity or cross-enhancement between influenza and RSV changes from year to year. The Discussion section elaborates on why this is expected, given the circulation of different strains, and why these parameters do not necessarily vary smoothly from year to year. Nevertheless, some interesting patterns emerge. Evidence is assessed though the coverage of the 95\% posterior credible intervals for the parameters $\sigma _{1}$ and $\sigma _{2}$. Cross-immunity for influenza is detected in epidemic years 2004--05 and 2006--07, meaning that prior infection with the circulating RSV strain may have inhibited infection with influenza, though there is little evidence of this effect in the opposite direction. In Figure~\ref{fig:real-data-posterior-sample-paths-disaggregated-all-years} this effect manifests in the lack of a defined peak in influenza infections following the peak in RSV infections. These relatively small influenza outbreaks may be insufficient to determine influenza's cross-interference effect for RSV, as evidenced by the fact that, in both years, $\sigma _{2}$ has approximately equal posterior probability of lying on either side of 1. Evidence of cross-enhancement for influenza, due to a prior infection with RSV, was found in epidemic years 2005--06, 2007--08, and 2008--09. Additionally, evidence for cross-enhancement for RSV was found in 2007-08, and 2008--09. However, results were inconclusive for the cross-interference outcome for RSV in 2005--06. Finally, epidemic season 2003--04 shows evidence of cross-enhancement for RSV from a prior influenza infection, but the magnitude of the effect in the other direction is inconclusive. This is expected, as this is the only year where the influenza peak preceded the peak in RSV and both peaks were well separated, resulting in a lack of RSV cases early in the season to inform the cross-interference parameter.

Contact transmission rates describe the degree of infectiousness of a given ARI strain. Their estimates and credible intervals are shown in the right panel of Figure~\ref{fig:real-data-posterior-samples-beta-sigma}. We can additionally investigate the relative sizes of the transmission rates for the particular strains of influenza and RSV in circulation. Assessment of this relationship is based on posterior probability of the difference between $\beta _{1}$ and $\beta _{2}$. We find evidence that RSV is more contagious than influenza in all years except 2003--04 and 2005--06, where evidence suggests a more infectious influenza strain. This pattern is also clearly discernible in Figure~\ref{fig:real-data-posterior-sample-paths-disaggregated-all-years}, where 2003--04 and 2005--06 are the only years studied where an influenza peak precedes a peak in RSV infections.

The importance of enabling the estimation of cross-interference and transmission rates and, consequently, disaggregating pathogen dynamics is highlighted in the Discussion section. Our analysis suggests that novel sentinel programs, like the one implemented at Hospital Central, provide a powerful tool to jointly estimate features of these epidemics.

\subsection{Model Assessment}

The model fit was assessed by analyzing predictive residuals and interpreting any deviations from their theoretical distribution. Results of the residual analysis are shown in Figures~3 through 5 of the Supplementary Material. A posterior sample of 100 step-ahead predicted states were transformed, via the observation process (i.e., rescaled and aggregated), and subtracted from the aggregated ARI reports data to obtain a sample from the residual posterior distribution (Figure~3 in the Supplementary Material, gray circles). The mean residuals are roughly centered around zero and are approximately normally distributed, as illustrated in Figure~4 in the Supplementary Material. However, some evidence of a pattern in the residuals warrants further investigation. For all six years there is a noticeable peak or drop of the residuals around the 20th week of the epidemic season. This again suggests that the two-pathogen SKM is not flexible enough to produce large epidemic peaks. The 20th week of the season coincides with the winter holidays and subsequent return of children to school. The increased social contact during this time violates the assumption of a constantly well-mixed population in the SKM which likely accounts for this model discrepancy. The background term is intentionally not flexible enough to capture these peaks to avoid overfitting the data. Indeed, empirical autocorrelation plots for the mean residuals, shown in Figure~5 in the Supplementary Material, detect statistically significant correlations over small lags which are not captured by the background process. The posterior variance of the predictive residuals is much greater during 2008--09 than the previous years, and the pattern in the residuals is more pronounced, including a very large peak around week 40 of the epidemic season. This spike corresponds directly to the the timing of the emergence of the influenza A (H1N1)pdm09 strain of influenza, in addition to the strain already in circulation. As expected, our parameteric model, which explicitly accounts for only two pathogens, is not able to capture the dynamics of an additional major outbreak. The Discussion section outlines ways in which these issues can be accounted for in the model, subject to the availability of appropriate data.

\section{Discussion}\label{sec:discussion}

We have presented an analysis of aggregated ARI report data and virological test data collected in the state of San Luis Potos\'{i}, M\'{e}xico. A Bayesian model was constructed based on a large-population (LNA) approximation to a stochastic kinetic SIR model for the dynamics for two ARI pathogens, influenza and RSV. In order to integrate multiple data sources and an empirical discrepancy model representing infections with other pathogens, we extended the existing marginal sampling approach based on the LNA. The inclusion of the additional virological data allowed us to jointly model disease dynamics of influenza and RSV in San Luis Potos\'{i}, to recover cross-interference parameters, and to disaggregate the trajectories of individual pathogens. The challenge of sampling from the posterior distribution over the model parameters was addressed by employing a population MCMC algorithm. Overall, our approach is general and can be applied to the problem of inference on other stochastic kinetic models as well as models of more than two interacting epidemics, when additional data sources are available to recover their individual dynamics.

\subsection{Joint inference and cross-interference}
\label{subsec:jointinf}

Joint modeling of the two pathogens enables estimation of cross-interference parameters and related qualitative features, such as the order and location of epidemic peaks of influenza and RSV in a given epidemic year. This is important because of the central role that cross-interference plays in shaping the evolutionary and epidemiological dynamics of multistrain pathogens (\cite{gjini2016direct,zhang2013co}). Changes in cross-interference are known to give rise to complex temporal dynamics in disease incidence and create oscillating time series from which it is difficult to estimate parameters that govern the underlying epidemiological processes (\cite{bhattacharyya2015cross,reich2013interactions}).

Studies about influenza-RSV interaction are predominantly qualitative (\cite{bhattacharyya2015cross,grondahl20142009,pinky2016coinfections}). Though cross-immunities have been estimated for dengue (\cite{reich2013interactions}) and influenza and pneumococcal pneumonia (\cite{shrestha2013identifying}), to our knowledge, there have been no quantitative studies of cross-interference between influenza and RSV. Furthermore, to our knowledge, our work is the first to estimate cross-interference parameters from an SKM.

Pathogen fitness is determined by processes that occur both within and between hosts (\cite{bashey2015within,mideo2008linking,van1995dynamics}). Many factors can drive cross-interference patterns between these viruses, such as vaccination programs (\cite{martcheva2007vaccine,rohani1999opposite}) and geographical location (\cite{bloom2013latitudinal}). Furthermore, the strain that is most successful within a host is not necessarily the one that can best be transmitted between hosts (\cite{mideo2008linking}). \textit{In-vitro} within-host interaction studies (\cite{mideo2008linking}) suggest that in cells coinfected with both pathogens, influenza blocks RSV infection (\cite{shinjoh2000vitro}). Observational studies by \cite{Anestad1987,miguez2016temporal,nishimura2005clinical} suggest a similar pattern. We note that our results do not contradict these studies, since we model between- rather than within-host interactions, focusing on the disease spread at the level of the host population.

\subsection{Prediction}
\label{subsec:prediction}

The model formulated in Section~\ref{sec:Modeling}, and the extended marginal sampling approach presented in Section~\ref{subsec:extendedpma} can, in principle, be applied over time domains of any length. However, in the context of our motivating problem, estimating the model for each epidemic year separately is consistent with the understanding of the way that different strains of influenza and RSV circulate from year to year. Parameters defining the circulation and interaction dynamics of influenza and RSV differ from strain to strain of the virus, and different strains of each virus circulate in different years. Specifically, RSV is classified into two major groups, A and B, each of which contains multiple variants (\cite{anderson1985antigenic,mufson1988respiratory,venter2001genetic}). Of the five years studied by \cite{peret1998circulation}, each year saw a shift in the predominant genotype or subtype of RSV so that no single genotype or subtype was dominant for more than one of the five years. \cite{peret1998circulation} hypothesized that newly introduced strains are better able to evade previously induced immunity in the population and, consequently, either circulate more efficiently or are more pathogenic. Similarly, influenza viruses continually change, due to antigenic drift, which requires vaccine reformulation before each annual epidemic (\cite{ferguson2003ecological}) and antigenic shift (\cite{CDC,rambaut2008genomic,zambon2001contribution}) which can lead to the emergence of new influenza epidemics. All of these effects combine to create unique conditions in each epidemic year. We note that sometimes such changes can occur within a single epidemic year, such as with the emergence of the global influenza A (H1N1)pdm09 pandemic in April 2009. Indeed, we observed that the model fitted using data from the epidemic year 2008--2009 was only partially able to capture the dynamics of this second epidemic wave of influenza in the latter part of the year.

\subsection{Model building}
\label{subsec:modelbuilding}

In future work we plan to address several questions. First, we believe that accounting for vaccination effects and seasonal forcing will yield additional insight into the interaction of these pathogens. Specifically, one important seasonal factor identified was the possible effect of holidays and school schedules on the system dynamics. This factor is difficult to incorporate directly into a SKM which assumes a well-mixed population. We hypothesize that this additional stochasticity may be introduced through a discrepancy process within the system equations, defining the LNA. Second, it may be realistic in some years to model multiple strains of influenza and RSV. Indeed, the emergence of additional strains, such as in April of 2009, is not suitably captured by our model in which parameters are essentially shared between strains of a single pathogen within each year. Therefore, modeling such strains as separate pathogens may yield better understanding of the circulation of ARI in the state. In order to fit these models, it will be critical to obtain new sources of data, such as vaccination reports, local climate variables, or virological tests for new strains, to identify additional model components.

\subsection{Conclusion}
\label{subsec:mexicandata}

The ability to understand the dynamics of infectious diseases is crucial for timely assessment of epidemics (\cite{Abat2016}). While infectious disease outbreaks are usually caused by a specific pathogen, symptoms caused by respiratory viruses are nonspecific, and do not allow for accurate diagnosis of the etiology of illness (including ARI in general as well as pneumonia). Therefore, virological and bacteriological methods are required to identify the agent responsible for a patient's disease. However, laboratory testing methods are not always available, may be costly, and, in some instances, have low sensitivity or specificity. As a result, laboratory evaluation of all respiratory infections is not feasible, and the role of specific pathogens cannot be fully ascertained. These limitations are also present when assessing epidemics at a population level. Although seasonal ARI epidemics tend to be dominated by one virus (such as influenza), the contribution of a second pathogen (such as RSV) needs to be considered when analyzing aggregated data (\cite{Charu2011}). Failing to consider these dynamics may result in overestimation of the effects of a specific virus as well as biased parameter estimates when modeling epidemics. Unfortunately, systematic data describing the epidemiological behavior during several consecutive seasons for some viruses, such as RSV, is not available in many countries (\cite{Simonsen2013}). In the present study we used influenza and RSV data, collected systematically and prospectively during a six-year period, together with ARI reports that are routinely collected by health-care authorities; this allowed us to analyze the contribution of these two pathogens to ARI epidemics. While influenza surveillance data is increasingly available worldwide, systematically obtained year-round RSV virological data from middle- and low-income countries is scarce, and no such data is currently available for M\'{e}xico (aside that from San Luis Potos\'{i}). Overall, our study underscores the relevance of considering more than one pathogen when assessing ARI epidemics. The availability of two-pathogen models could have important applications not only during interpandemic winter seasons but also during the emergence of pandemic viruses, such as influenza A (H1N1)pdm09 virus and SARS-CoV-2.

\begin{appendix}
\section*{Large-population approximation to SKM}\label{app:A1}
%
%
%
A large-population approximation of the CME (\ref{eqs:ME}) is given by the van Kampen expansion which can then be computed via the Linear Noise Approximation (LNA) (\cite{van1992stochastic}). For large $\Omega $ the system states $X$ can be expressed as the sum of a deterministic term $\phi : [0, T]\to \mathbb{R}^{+\dim (X)}$ and a stochastic term $\xi $,
%
\begin{equation}
\label{eqs:_LNA} X(t) = \Omega \phi (t) + \Omega ^{1/2} \xi (t), \quad t
\in [0, T].
\end{equation}
Assuming constant average concentration, the magnitude of the stochastic component will increase as the square root of population size. Let $V =(v_{1},\ldots ,v_{\mathcal{R}})$ be a $\dim (X) \times \mathcal{R}$ stoichiometric matrix that describes changes in the population size, due to each of the $\mathcal{R}$ reactions. The time evolution of the term of order $\Omega ^{1/2}$ (\cite{van1992stochastic}), $\phi _{i}(t)=\lim_{\Omega , X \longrightarrow \infty }X_{i}(t)/ \Omega $, is governed by the ODE initial value problem,
%
\begin{equation}
\begin{cases} \frac{d\phi _{i} (t)}{dt} =\sum_{j=1}^{\mathcal{R}}
V_{ij}a_{j}\bigl( \phi (t)\bigr), & t\in (0,T], i=1,\ldots ,
\dim (X),
\\
\phi _{i} (0) = \phi _{0}, & i=1,\ldots , \dim (X),
\label{eqs:_detpartODE} \end{cases}
\end{equation}
where $\phi _{0} = X(0)/\Omega $. The full expressions for the macroscopic equations for the two-pathogen model in the first line of (\ref{eqs:_detpartODE}) are
%
\begin{equation}
\begin{aligned} \frac{d\phi _{1} (t)}{dt} &= \mu - \beta _{2}
\lambda _{2} \phi _{1}-\beta _{1}\lambda
_{1}\phi _{1}-\mu \phi _{1},
\\
\frac{d\phi _{2} (t)}{dt} &= \beta _{1}\lambda _{1} \phi
_{1}- \gamma \phi _{2}-\mu \phi _{2},
\\
\frac{d\phi _{3} (t)}{dt} &= \gamma \phi _{2}-{\sigma _{2}}\beta
_{2} \lambda _{2} \phi _{3}-\mu \phi
_{3},
\\
\frac{d\phi _{4} (t)}{dt} &= \beta _{2}\lambda _{2} \phi
_{1}- \gamma \phi _{4}-\mu \phi _{4},
\\
\frac{d\phi _{5} (t)}{dt} &= {\sigma _{2}}\beta _{2}\lambda
_{2} \phi _{3}-\gamma \phi _{5}-\mu \phi
_{5},
\\
\frac{d\phi _{6} (t)}{dt} &= \gamma \phi _{4}-\mu \phi _{6}-{\sigma
_{1}} \beta _{1}\lambda _{1} \phi
_{6},
\\
\frac{d\phi _{7} (t)}{dt} &= -\gamma \phi _{7}+{\sigma _{1}} \beta
_{1} \lambda _{1} \phi _{6}-\mu \phi
_{7},
\\
\frac{d\phi _{8} (t)}{dt} &= \gamma \phi _{7} + \gamma \phi _{5}-
\mu \phi _{8}. \end{aligned} %
\label{eq:ODE3}
\end{equation}
%
%
%
\begin{sidewaystable}
\resizebox{1 \textwidth}{!}{  
\begin{tabular}{c}
$A(t)$ \\
$\begin{bmatrix}
-(\beta _{2}\lambda _{2}+\beta _{1}\lambda _{1}+\mu ) & -\beta _{1}\phi _{1} & 0 & -\beta _{2}\phi _{1}& -\beta _{2}\phi _{1}&0&-\beta _{1}\phi _{1}&0\\
\beta _{1}\lambda _{1} & \beta _{1}\phi _{1}-(\gamma +\mu )& 0 & 0 & 0 & 0 & \beta _{1}\phi _{1}& 0\\
0 & \gamma & -(\beta _{2}\lambda _{2}{\sigma _{2}}+\mu ) & -\beta _{2}\phi _{3}{\sigma _{2}} & -\beta _{2}\phi _{3}{\sigma _{2}} &0 &0 &0\\
\beta _{2}\lambda _{2} & 0 & 0 & \beta _{2}\phi _{1}-(\gamma +\mu ) & \beta _{2}\phi _{1} & 0 & 0 & 0 \\
0 & 0 & \beta _{2}\lambda _{2}{\sigma _{2}} & \beta _{2}\sigma \phi _{3} & \beta _{2}{\sigma _{2}}\phi _{3}-(\gamma +\mu )&0&0&0\\
0 & -\beta _{1}\phi _{6}{\sigma _{1}} & 0 & \gamma & 0 & -\beta _{1}\lambda _{1}{\sigma _{1}}-\mu & -\beta _{1}{\sigma _{1}}\phi _{6} &0\\
0 & \beta _{1}{\sigma _{1}}\phi _{6}& 0 & 0 & 0 & \beta _{1}\lambda _{1}{\sigma _{1}} & \beta _{1}{\sigma _{1}}\phi _{6}-(\gamma +\mu )&0\\
0 & 0& 0& 0& \gamma & 0 & \gamma & -\mu
\end{bmatrix}$
\vspace*{20pt}\\
$B(t)$ \\
$\begin{bmatrix}
\beta _{2}\lambda _{2}\phi _{1} & & & & & & & \\
\beta _{1}\lambda _{1}\phi _{1} & -\beta _{1}\phi _{1}\lambda _{1} & & -\beta _{2}\lambda _{2}\phi _{1}&0&0&0&0\\
\phi _{1}\mu +\mu & & & & & & &\\
-\beta _{1}\lambda _{1}\phi _{1} & \beta _{1}\lambda _{1}\phi _{1}+\phi _{2}(\gamma +\mu ) & -\gamma \phi _{2} & 0 &0 &0 &0 &0\\
0 & -\gamma \phi _{2} & \gamma \phi _{2} & & & & &\\
& & \beta _{2}\phi _{3}\lambda _{2}{\sigma _{2}}+\mu \phi _{3}& 0&-\beta _{2}\phi _{3}\lambda _{2}{\sigma _{2}} & 0 & 0 &0 \\
-\beta _{2}\phi _{1}\lambda _{2} & 0 & 0 & \beta _{2}\phi _{1}\lambda _{2}+\phi _{4}(\gamma +\mu ) & 0 & -\gamma \phi _{4} &0 &0\\
& & & & \phi _{5}(\gamma +\mu ) & & &\\
0 & 0 & -\beta _{2}\lambda _{2}{\sigma _{2}} \phi _{3} & & \beta _{2}\lambda _{2}{\sigma _{2}}\phi _{3} & 0 & 0 & -\gamma \phi _{5}\\
0 & 0 & 0 & -\gamma \phi _{4} & 0 & \gamma \phi _{4} + \mu \phi _{6} & & \\
& & & & &\beta _{1}\lambda _{1}{\sigma _{1}}\phi _{6}& -\beta _{1}\phi _{6}\lambda _{1}{\sigma _{1}}& 0 \\
0 & 0 & 0 & 0 & 0 & -\beta _{1}{\sigma _{1}}\lambda _{1}\phi _{6} & \beta _{1}\lambda _{1}{\sigma _{1}}\phi _{6} & \\
& & & & & & (\gamma +\mu )\phi _{7} & -\gamma \phi _{7}\\
0 & 0 & 0 & 0 & -\gamma \phi _{5} & 0 & -\gamma \phi _{7} & \gamma \phi _{5} \\
& & & & & & & \gamma \phi _{7} + \mu \phi _{8}
\end{bmatrix}$\\
\end{tabular}}
\caption{Matrices defining the diffusion approximation of the master equation}
\label{table:matricesAandB}
\end{sidewaystable}
\\
\\
The stochastic process $\xi $ is governed by the It\^{o} diffusion equation,
%
\begin{equation}
d\xi (t) = A(t)\xi (t)dt + \sqrt{B(t)}dW(t), \quad t\in [0, T], \label{eqs:langevin}
\end{equation}
where $A(t) = \partial V   a(\phi (t))/\partial \phi (t)$, $B(t) = V \operatorname{diag}\{a(\phi (t))\}  V^{\top }$, and $W(t)$ denotes the $\dim (X)$-dimensional Wiener process (\cite{gillespie2007stochastic,van1992stochastic}). Expressions for both matrices in the two-pathogen model are provided in Table~\ref{table:matricesAandB}. For fixed or Gaussian initial conditions, the SDE in (\ref{eqs:langevin}) can be solved analytically. The solution is a Gaussian process with mean $\tilde{\xi }$ and covariance $C$ (\cite{van1992stochastic}), that is,
%
\begin{equation}
\label{eqs:NoiseDistrib} \xi (t) \sim {\mathcal{N}} \bigl(\tilde{\xi } (t ), C(t) \bigr),
\quad t\in [0, T],
\end{equation}
where $\tilde{\xi }(t)$ and $C(t)$ are obtained (see
\citeauthor{van1992stochastic} (\citeyear{van1992stochastic}, pp.~244--258)) by solving the ODE initial value
problem,
%
\begin{equation}
\begin{cases} \frac{\partial \tilde{\xi }(t)}{\partial t} = \Phi (t) \tilde{\xi }(t), & t\in
(0, T],
\\
\xi (0) = \tilde{\xi }_{0}, \end{cases} %
\end{equation}
where $\Phi (t)$ is the evolution or fundamental matrix (\cite{grimshaw1991nonlinear}) determined by the matrix equation,
%
\begin{equation}
\begin{cases} \dot{\Phi }(t)=A(t)\Phi (t),& t\in (0, T],
\\
\Phi (0) = I. \end{cases}
\end{equation}
A standard choice (e.g., \cite{fearnhead2014inference}) is to set
$\tilde{\xi }(0)=0$, from which it follows that $\tilde{\xi }(t)=0$ for all
$t\in [0, T]$. The covariance $C$ is obtained by solving
%
\begin{equation}
\begin{cases} \frac{dC(t)}{dt} = C(t)A(t)^{T}+A(t)C(t)+B(t),&
t\in (0, T],
\\
C(0) = C_{0}. \end{cases} %
\label{eqn:SDEterm}
\end{equation}
It follows from (\ref{eqs:_LNA}) and (\ref{eqs:NoiseDistrib}) that the transition densities of $X(t)$ are given by (\ref{lik:process}).
\end{appendix}

\section*{Acknowledgements}
The authors thank the following people for helpful comments and suggestions: Grzegorz A. Rempala (Mathematical Biosciences Institute, The Ohio State University) and Leticia Ramirez (Centro de Investigaci\'{o}n en Matem\'{a}ticas). The authors also wish to thank The Ohio State University and Centro de Investigaci\'{o}n en Matem\'{a}ticas (CIMAT) for making this collaboration possible. Finally, we thank the anonymous reviewers and Associate Editor for invaluable comments and suggestions.

\section*{Funding}
This research was supported in part by the Mathematical Biosciences Institute (MBI) and the National Science Foundation under grant DMS 1440386.

\newpage\newpage\newpage\newpage\newpage\newpage\newpage\newpage
\bibliographystyle{imsart-nameyear} 
\bibliography{BI_references}       



\end{document}


\begin{frontmatter}

\title{Supplement to ``Inference for stochastic kinetic models from multiple data sources for joint estimation of infection dynamics from aggregate reports and virological data''}
\runtitle{Inference for SKMs from multiple data sources}


\author{\fnms{Oksana A.} \snm{Chkrebtii}\thanksref{t1,t2}\ead[label=e1]{oksana@stat.osu.edu}},
\author{\fnms{Yury E.} \snm{Garc\'ia}\thanksref{t1,t3}\ead[label=e1]{yury@cimat.mx}},
\author{\fnms{Marcos A.} \snm{Capistr\'an}\thanksref{t3}\ead[label=e1]{marcos@cimat.mx}},
\and
\author{\fnms{Daniel E.} \snm{Noyola}\thanksref{t4}\ead[label=e1]{dnoyola@uaslp.mx}}

\thankstext{t1}{Indicates equal contribution}
\thankstext{t2}{Department of Statistics, The Ohio State University, Columbus, Ohio, USA}
\thankstext{t3}{\'{A}rea de Matem\'{a}ticas B\'{a}sicas, Centro de Investigaci\'on en Matem\'aticas, Guanajuato, Gto., M\'exico}
\thankstext{t4}{Department of Microbiology, Faculty of Medicine, Universidad Aut\'onoma de San Luis Potos\'i, M\'exico}

\end{frontmatter}


\textbf{Additional Figures and Results}
The following are additional figures and results from the analysis of the data described in the paper. The analysis was implemented in Python.  Python module ``corner'' \citep{ForemanMackey2016} was used to display the bivariate posterior correlation plots.  
\\\\\\\\\\\\\\\\\\\
\begin{figure}
\centering
\includegraphics[width=\textwidth]{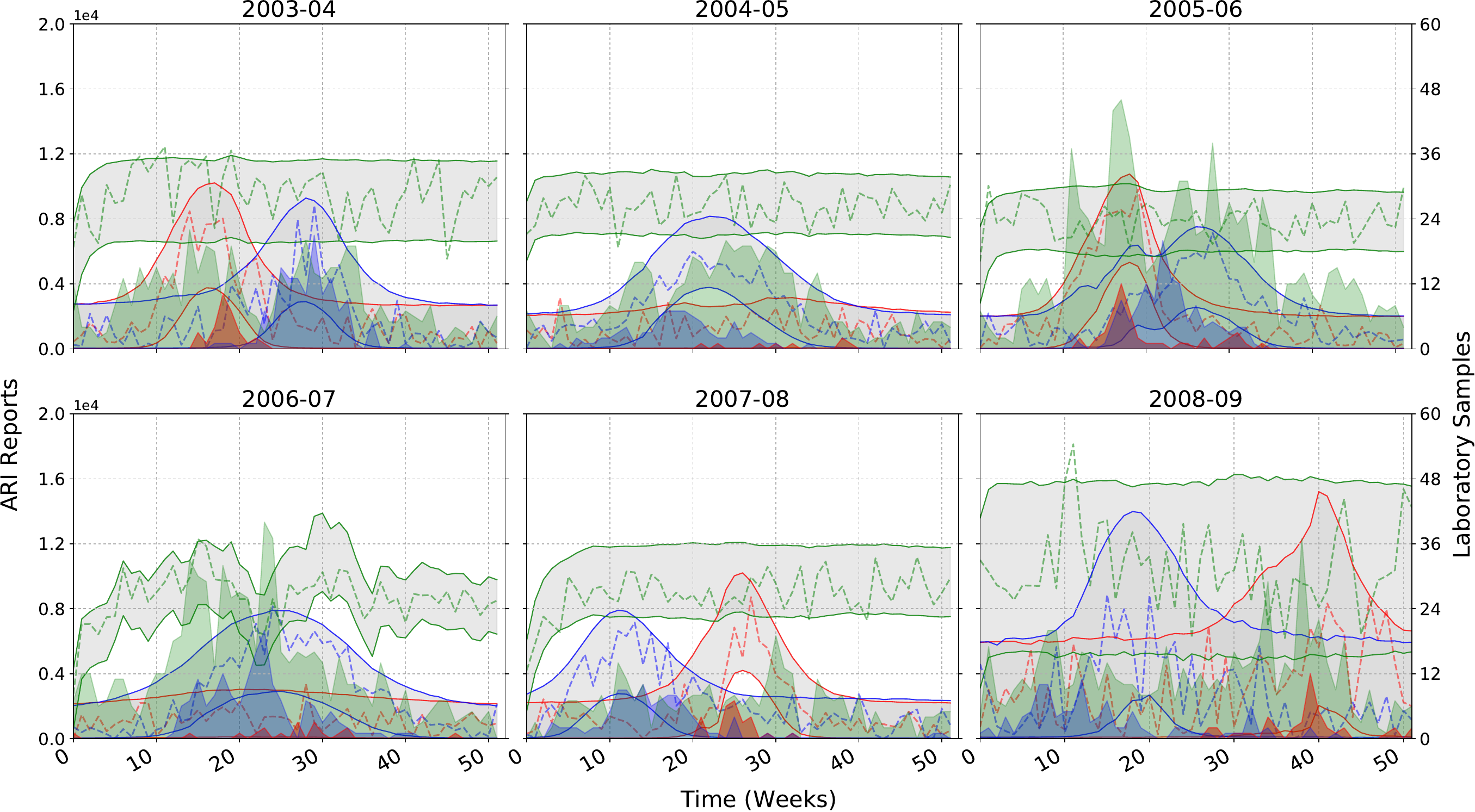}
\caption{Posterior predictive summaries for disaggregated reports based on data from San Luis Potos\'i, M\'exico.  Left axis: Grey bands represent 95\% prediction intervals for influenza (red outline), RSV (blue outline) and background infections (green outline).  Dashed lines represent maximum a-posteriori predictions.  Right axis: Reports of influenza and RSV from ARI affected children under 5 years of age at Hospital Central, with tests positive for influenza, RSV, and neither, are shown as shaded red, blue, and green polygons, respectively. }
\label{fig:real-data-postpred-sample-paths-disaggregated-all-years}
\end{figure}
\afterpage{%
\begin{figure}
\centering
\includegraphics[width=\textwidth]{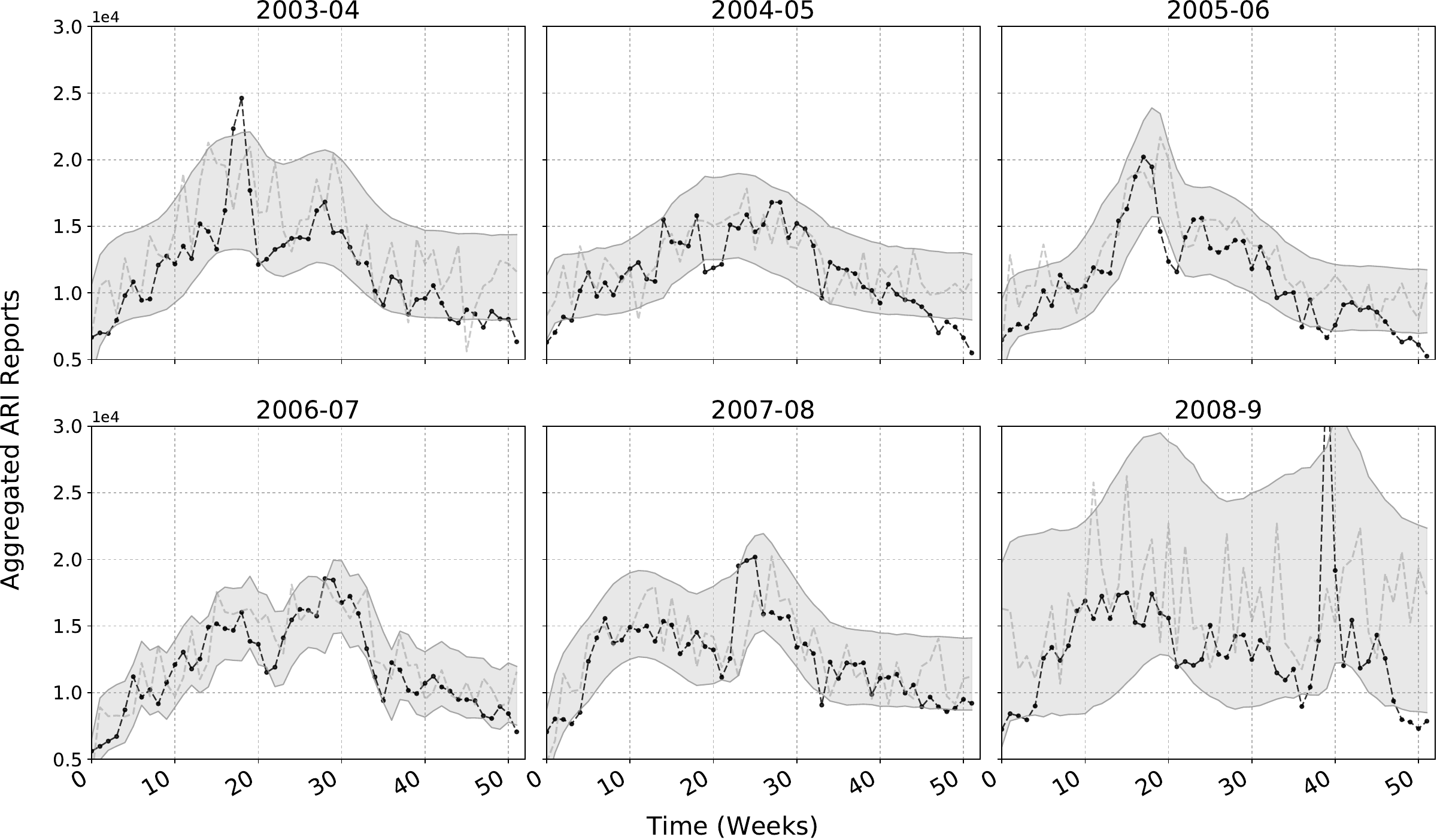}
\caption{ Posterior predictive summaries for aggregated ARI reports from San Luis Potos\'i, M\'exico.  The aggregated data, measured from August to July of the following year, is represented by black dots.  Gray bands represent point-wise 95\% prediction intervals and the gray dashed line shows a maximum a-posteriori predictions.}
\label{fig:real-data-postpred-sample-paths-all-years}
\end{figure}
}
\afterpage{%
\begin{figure}
\centering
\includegraphics[width=0.7\textwidth]{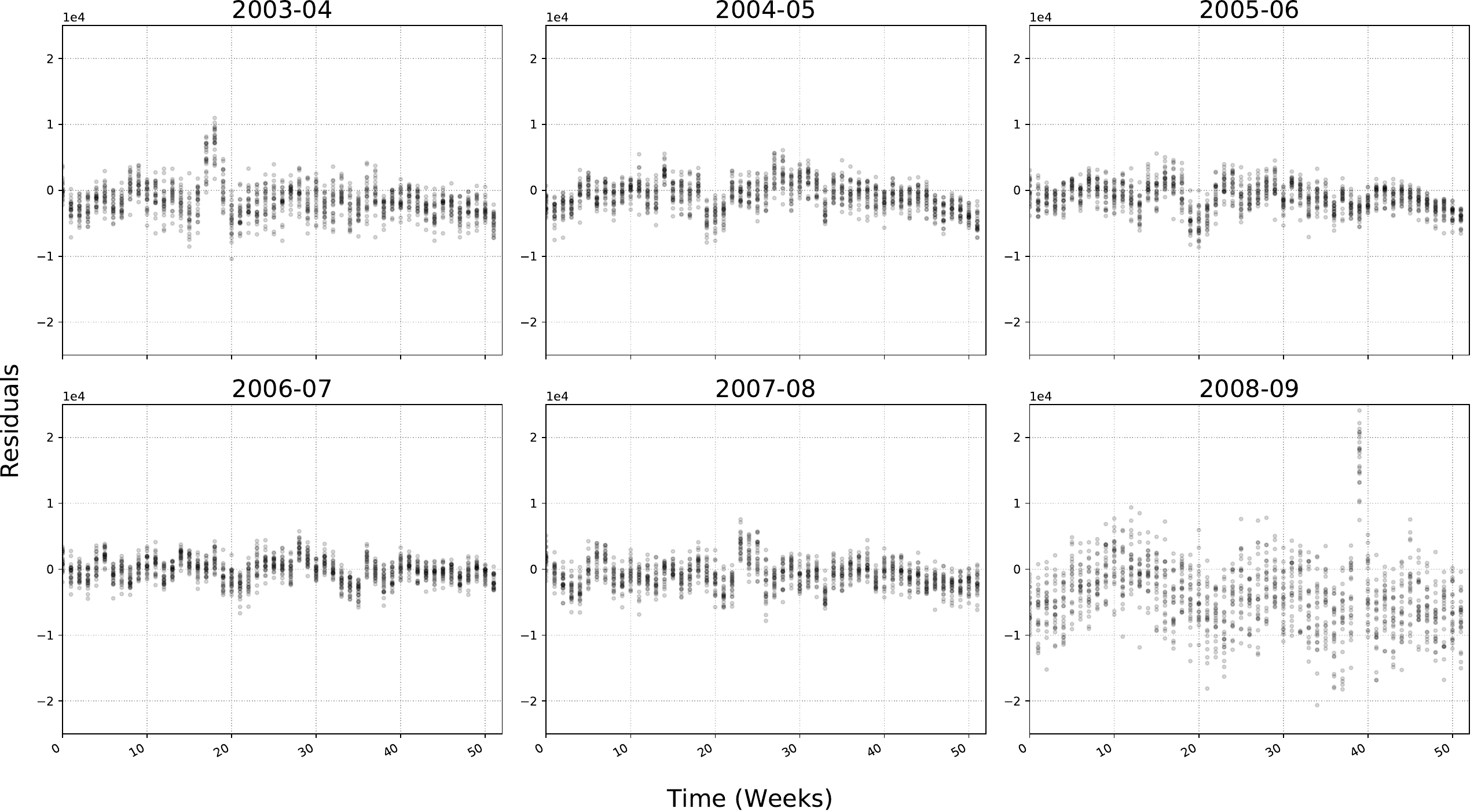}
\caption{Sample of 100 step-ahead residuals (gray circles) for the fitted model based on data from San Luis Potos\'i, M\'exico for each of six epidemic years. }
\label{fig:real-data-posteriorpredictive-residuals-all-years}
\end{figure}
\begin{figure}
\centering
\includegraphics[width=0.7\textwidth]{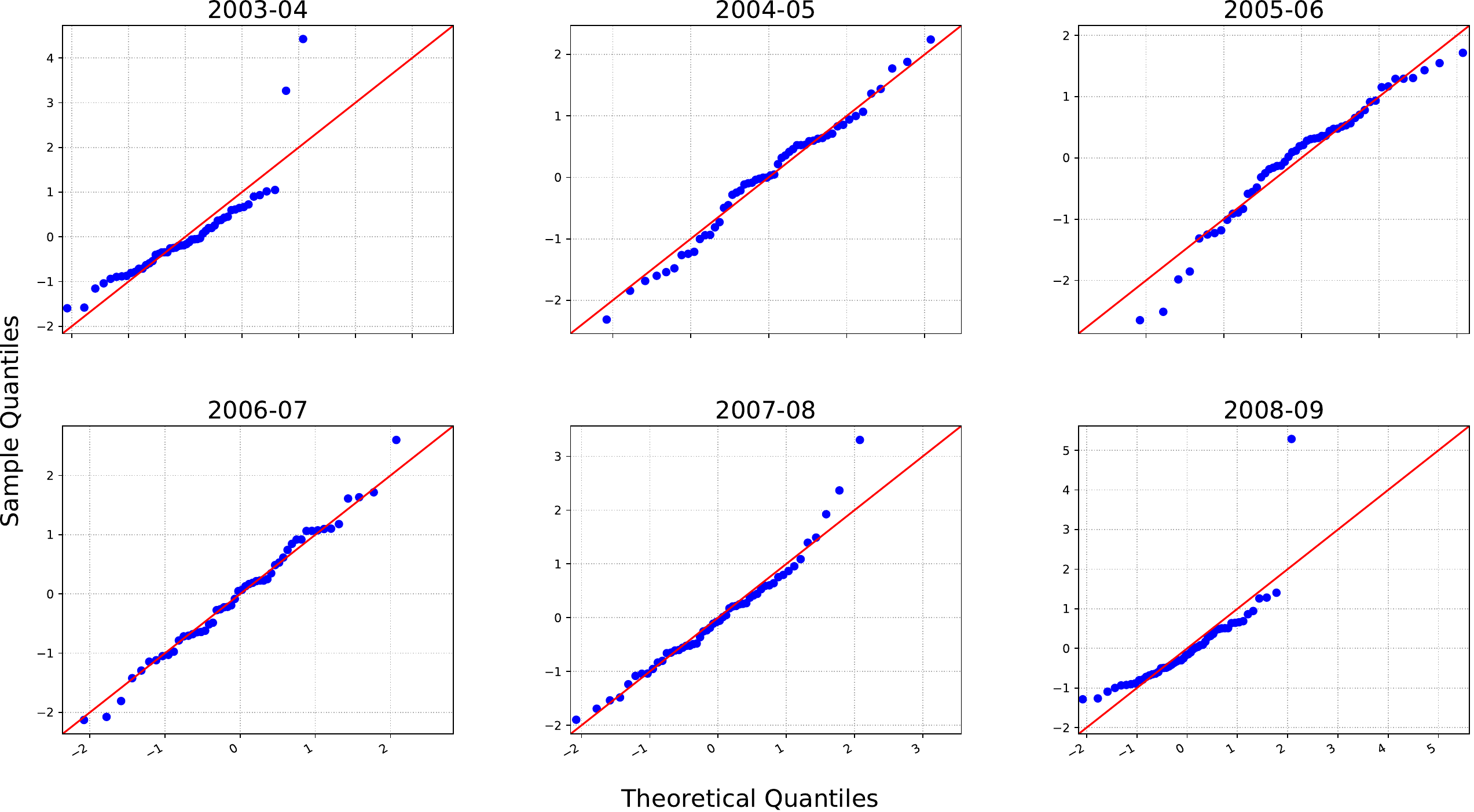}
\caption{Quantile-quantile plots for step-ahead residuals for the fitted model based on data from San Luis Potos\'i, M\'exico for each of six epidemic years.}
\label{fig:real-data-posteriorpredictive-qqplot-all-years}
\end{figure}
\begin{figure}
\centering
\includegraphics[width=0.7\textwidth]{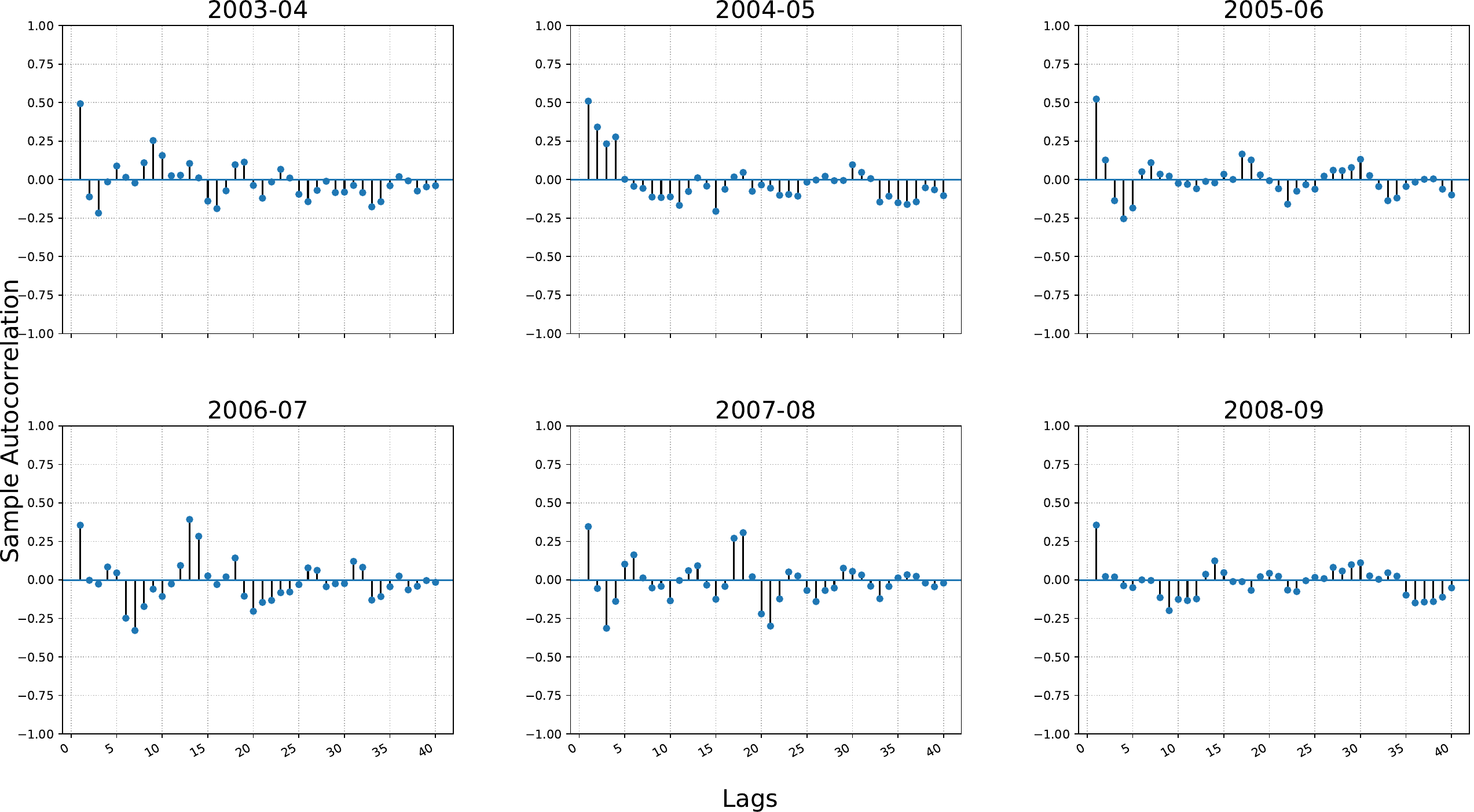}
\caption{Empirical autocorrelation plots for step-ahead residuals for the fitted model based on data from San Luis Potos\'i, M\'exico for each of six epidemic years.}
\label{fig:real-data-posteriorpredictive-acfplot-all-years}
\end{figure}
}
\afterpage{%
\begin{figure}
\includegraphics[scale = 0.15]{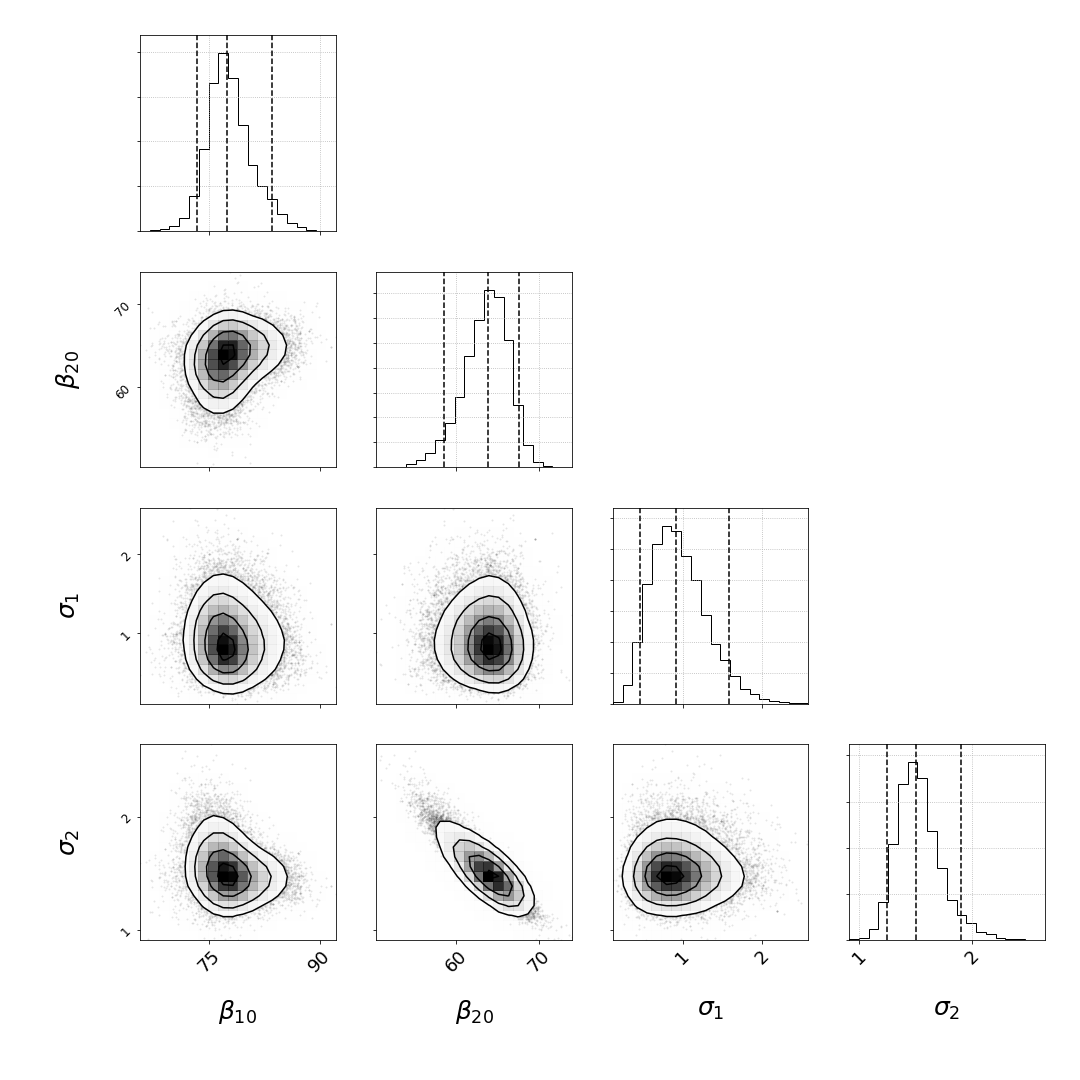}
\includegraphics[scale = 0.15]{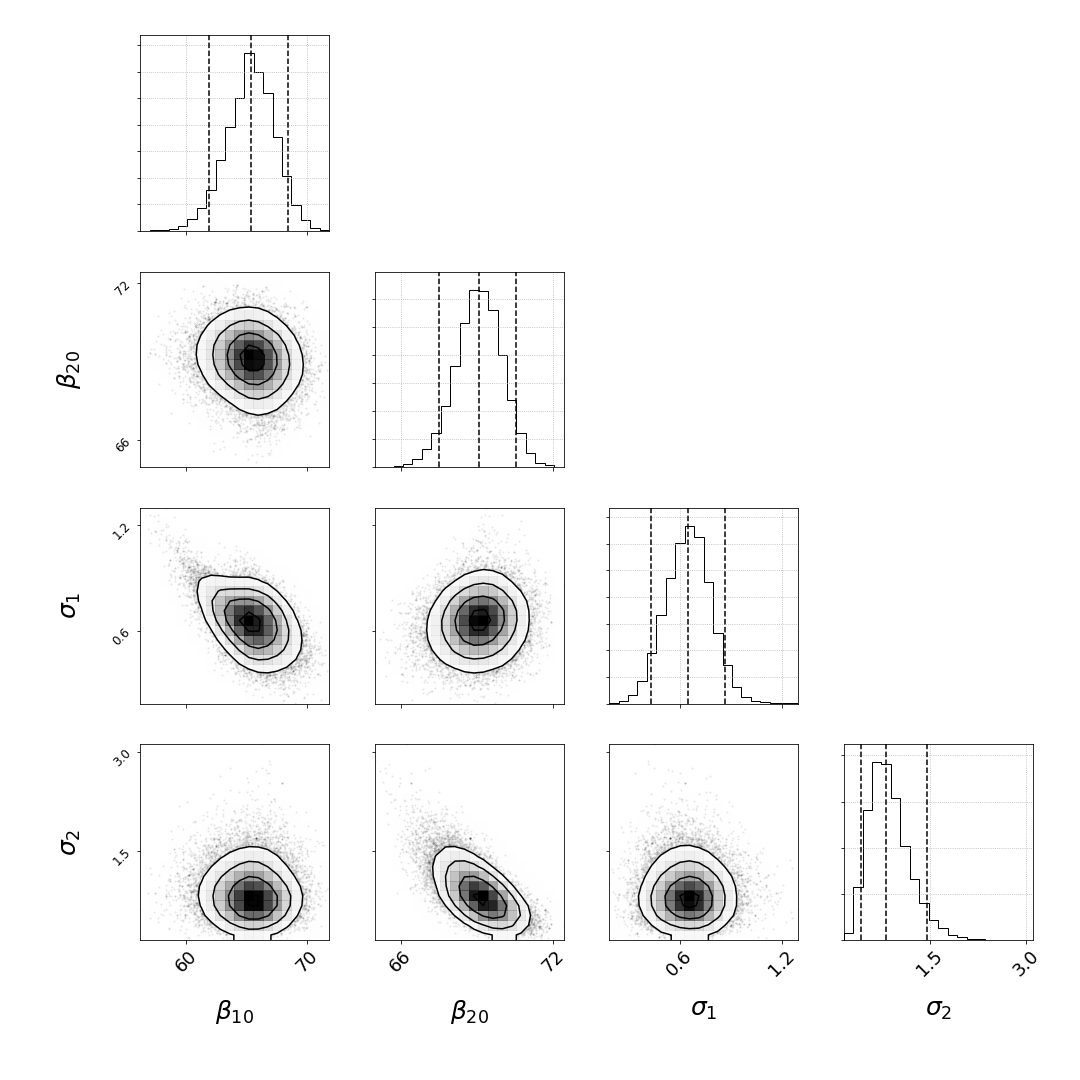}
\caption{Pairwise marginal contours for parameters defining the SKM based on data from San Luis Potos\'i, M\'exico for the years 2003-04 (left) and 2004-05 (right).}
\end{figure}
\begin{figure}
\includegraphics[scale = 0.15]{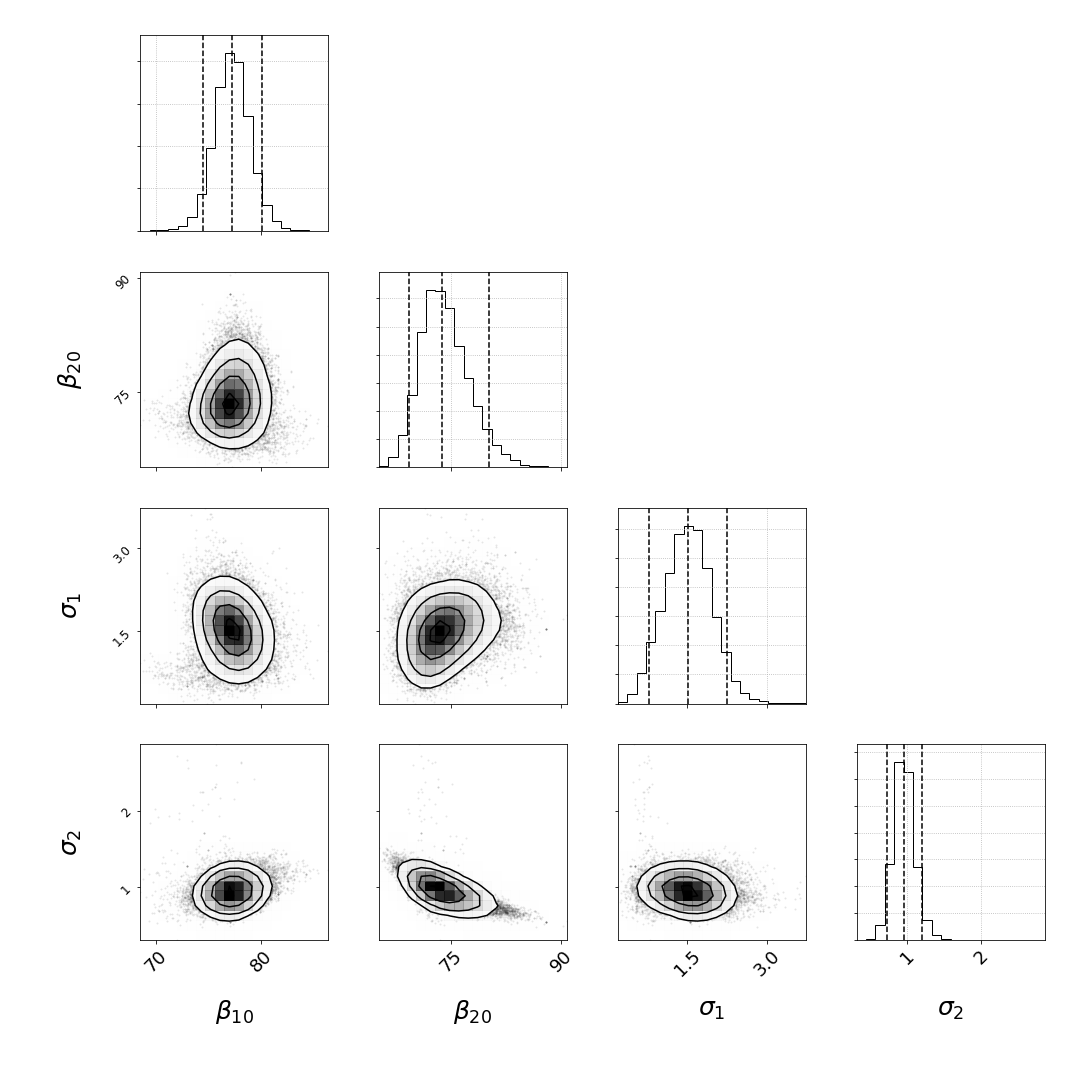}
\includegraphics[scale = 0.15]{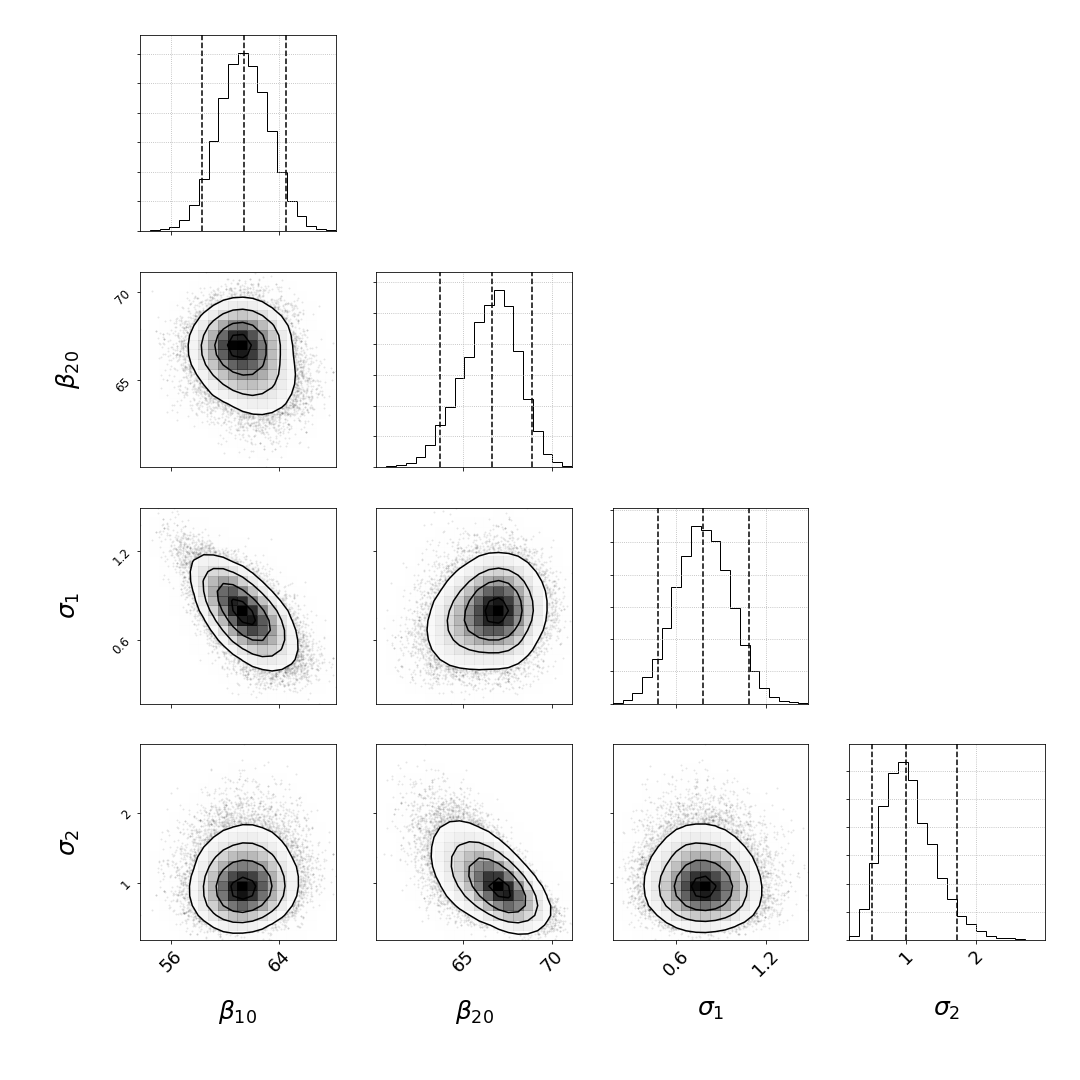}
\caption{Pairwise marginal contours for parameters defining the SKM based on data from San Luis Potos\'i, M\'exico for the years 2005-06 (left) and 2006-07 (right).}
\end{figure}
\begin{figure}
\includegraphics[scale = 0.15]{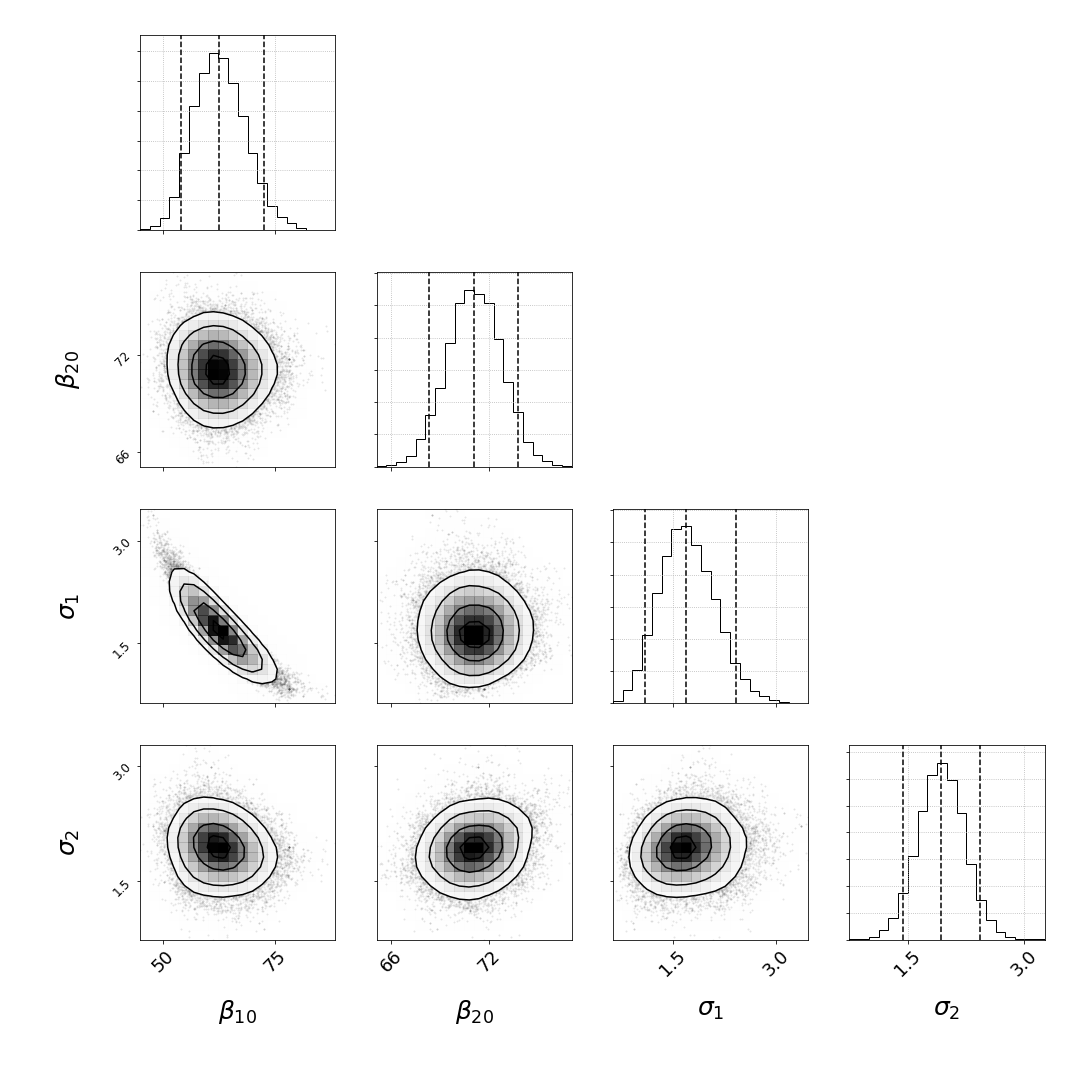}
\includegraphics[scale = 0.15]{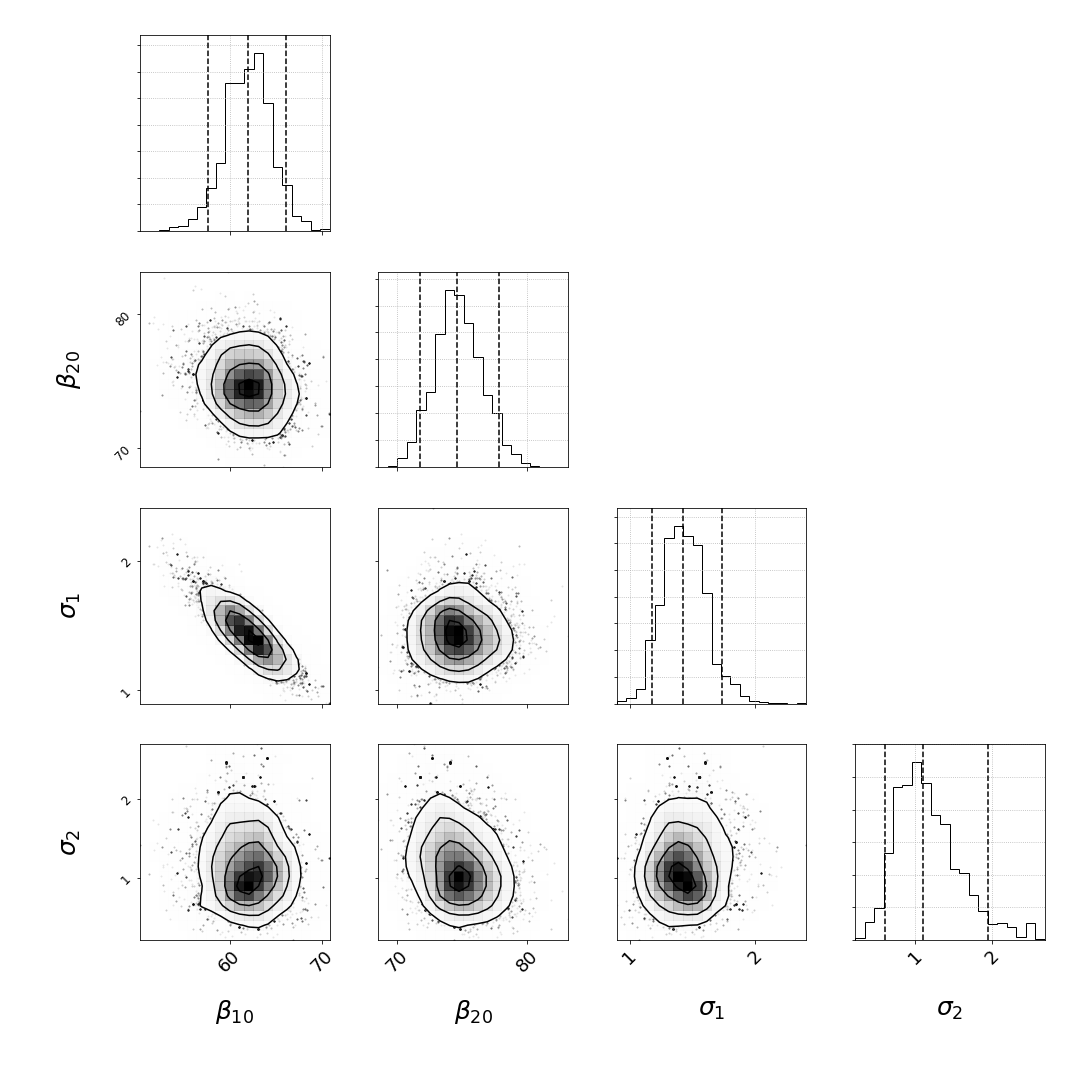}
\caption{Pairwise marginal contours for parameters defining the SKM based on data from San Luis Potos\'i, M\'exico for the years 2007-08 (left) and 2008-09 (right).}
\end{figure}
\clearpage
}%
\afterpage{%
\begin{table}
\centering
\caption{Posterior summaries for the year 2003-04}
\begin{tabular}{llllll}
\hline
Parameter  &  MAP    & $95\%$ Credible Intervals      & Parameter&   MAP    & $95\%$ Credible Intervals \\
\hline
$\beta_1$  &77.5040
               & (72.4580, 84.7378)    
&$X_{SS}$  & 2.2814e+06
      &(2.2788e+06	2.2829e+06)\\
$\beta_2$  & 63.8230
               & (57.3645, 68.1798)   
& $X_{IS}$  & 4.1446e+02
   &(2.2771e+01, 2.0284e+03)\\
$\sigma_1$   & 0.9126
               & (0.3912,	1.7299)       
&$X_{RS}$  & 4.1638e+04
     &(3.9394e+04,	4.5774e+04)\\       
$\sigma_2$   & 1.5060
               & (1.2079,	1.9936)       
&$X_{SI}$  & 7.1892e+02
     &(2.3636e+02,	2.3190e+03)\\       
$\Sigma$   & 6.3606e-07
	& (4.5055e-07,	8.6406e-07)
&  $X_{RI}$  & 3.0355e+02
 	&(1.0429e+01,	1.9849e+03)\\
$r$  & 7.5662e-02
               & (5.6196e-02,	9.6458e-02)       
&$X_{SR}$  & 2.7840e+04
   &(2.6845e+04,	2.9279e+04)\\
$c$        & 2.6297e-02
              & (1.7651e-02, 3.9532e-02)       
&$X_{IR}$  & 3.6219e+02
     &(2.9219e+01,	1.1983e+03)\\
$\nu$       &  4.5736e-01
                & (3.0047e-01,	6.0164e-01)     
&$X_{RR}$  & 1.4793e+05
     &(1.4278e+05,	1.4914e+05)\\
$v$        & 2.4594e-01
                & (1.6989e-01,	3.1062e-01)       &\\
\hline
\end{tabular}\label{tab:03-04}
\normalsize
\end{table}
\begin{table}
\centering
\caption{Posterior summaries for the year 2004-05}
\begin{tabular}{llllll}
\hline
Parameter  &  MAP    & $95\%$ Credible Intervals      & Parameter&   MAP    & $95\%$ Credible Intervals \\
\hline
$\beta_1$  &65.3822
               & (61.1196, 69.0546)    
&$X_{SS}$  & 2.2820e+06
      &(2.2818e+06,	2.2820e+06)\\
$\beta_2$  & 69.0853
               & (67.1501, 70.8143)   
& $X_{IS}$  & 2.3527e+02
   &(8.3438e+01,	3.2589e+02)\\
$\sigma_1$   & 0.6522
               & (0.3950,	0.9126)       
&$X_{RS}$  & 4.1924e+04
     &(4.1859e+04,	4.2601e+04)\\       
$\sigma_2$   & 0.8140
               & (0.3615,	1.6152)       
&$X_{SI}$  & 7.5687e+02
     &(1.9670e+02,	8.8676e+02)\\       
     $\Sigma$   & 2.6696e-07
	& (2.4274e-07,	3.5849e-07)
&  $X_{RI}$  & 4.7793e+02
 	&(1.2656e+02,	5.2752e+02)\\
$r$  & 1.3747e-01
               & (1.3363e-01,	1.5745e-01)       
&$X_{SR}$  & 2.7413e+04
   &(2.7358e+04, 2.7870e+04)\\
$c$        & 2.0883e-02
              & (1.8766e-02,	2.1947e-02)       
&$X_{IR}$  & 6.2315e+01
     &(6.4447e+00	1.3351e+02)\\
$\nu$       &  1.8568e-01
                & (1.7682e-01,	1.8698e-01)     
&$X_{RR}$  & 1.4715e+05
     &(1.4706e+05	1.4718e+05)\\
$v$        & 3.6186e-01
                & (3.3241e-01,	4.7776e-01)       &\\
\hline
\end{tabular}\label{tab:04-05}
\normalsize
\end{table}
\begin{table}
\centering
\caption{Posterior summaries for the year 2005-06}
\begin{tabular}{llllll}
\hline
Parameter  &  MAP    & $95\%$ Credible Intervals      & Parameter&   MAP   & $95\%$ Credible Intervals \\
\hline
$\beta_1$  &77.2244
               & (73.7990, 80.7387)    
&$X_{SS}$  & 2.2821e+06
      &(2.2818e+06,	2.2822e+06)\\
$\beta_2$  & 73.8391
               & (68.6161,	81.5908)   
& $X_{IS}$  & 2.8062e+02
   &(2.3144e+01,	5.1736e+02)\\
$\sigma_1$   & 1.5203
               & (0.6744,	2.3864)       
&$X_{RS}$  & 4.2429e+04
     &(4.2238e+04,	4.2440e+04)\\       
$\sigma_2$   & 0.9525
               & (0.6913,	1.2683)       
&$X_{SI}$  & 1.7266e+02
     &(9.6959e-01,	2.3065e+02)\\       
$\Sigma$   & 3.1787e-07
	& (1.8396e-07,	5.3983e-07)
&  $X_{RI}$  & 9.0837e+01
 	&(0.0000,	3.8530e+02)\\
$r$  & 8.1969e-02
               & (6.4701e-02,	9.9263e-02)       
&$X_{SR}$  & 2.7426e+04
   &(2.7386e+04,	2.7485e+04)\\
$c$        & 2.6065e-02
              & (1.7037e-02,	3.2837e-02)       
&$X_{IR}$  & 2.5274e+02
     &(4.2203e+01,	3.1079e+02)\\
$\nu$       &  3.1672e-01
                & (1.3699e-01,	5.5261e-01)     
&$X_{RR}$  & 1.4748e+05
     &(1.4733e+05,	1.4756e+05)\\
$v$        & 1.2203e-01
                & (1.0745e-01,	1.4894e-01)       &\\
\hline
\end{tabular}\label{tab:05-06}
\normalsize
\end{table}
\begin{table}
\centering
\caption{Posterior summaries for the year 2006-07}
\begin{tabular}{llllll}
\hline
Parameter  &  MAP    & $95\%$ Credible Intervals      & Parameter&   MAP    & $95\%$ Credible Intervals \\
\hline
$\beta_1$  & 61.3841
               & (57.7805,	65.1055)    
&$X_{SS}$  & 2.2818e+06
      &(2.2809e+06,	2.2828e+06)\\
$\beta_2$  & 66.6062
               & (63.2255,	69.2414)   
& $X_{IS}$  & 1.6434e+03
   &(2.9057e+02,	2.1287e+03)\\
$\sigma_1$   & 0.7784
               & (0.4201,	1.1397)       
&$X_{RS}$  & 4.1571e+04
     &(4.0062e+04,	4.2220e+04)\\       
$\sigma_2$   & 0.9984
               & (0.4436	1.8937)       
&$X_{SI}$  & 4.0613e+02
     &(1.0317e+02,	9.0607e+02)\\         
$\Sigma$   & 2.3968e-08
	& (1.0634e-09,	9.8463e-08)
&  $X_{RI}$  & 4.7257e+02
 	&(9.9385e+00,	2.4886e+03)\\
$r$  & 1.8967e-01
               & (1.5755e-01,	2.3205e-01)       
&$X_{SR}$  & 2.8478e+04
   &(2.6314e+04, 2.9670e+04)\\
$c$        & 6.5500e-03
              & (3.2692e-03, 9.9999e-03)       
&$X_{IR}$  & 3.7822e+02
     &(1.1458e+01,	9.5487e+02)\\
$\nu$       &  6.5254e-01
                & (4.6226e-01,	8.2090e-01)     
&$X_{RR}$  & 1.4532e+05
     &(1.4391e+05,	1.4802e+05)\\
$v$        & 1.9743e-01
                & (1.9743e-01,	1.9743e-01)       &\\
\hline
\end{tabular}\label{tab:06-07}
\normalsize
\end{table}
\clearpage
}
\afterpage{%
\begin{table}
\centering
\caption{Posterior summaries for the year 2007-08}
\begin{tabular}{llllll}
\hline
Parameter  &  MAP    & $95\%$ Credible Intervals      & Parameter&   MAP    & $95\%$ Credible Intervals \\
\hline
$\beta_1$  & 62.4283
               & (52.6160,	74.4917)    
&$X_{SS}$  & 2.2727e+06
      &(2.2691e+06,	2.2749e+06)\\
$\beta_2$  & 71.1031
               & (67.8569,	74.3190)   
& $X_{IS}$  & 2.8004e+02
   &(8.2947e+00,	1.2076e+03)\\
$\sigma_1$   & 1.6955
               & (0.9928	2.5656)       
&$X_{RS}$  & 4.3510e+04
     &(3.9274e+04,	4.6981e+04)\\       
$\sigma_2$   & 1.9303
               & (1.3325,	2.5360)       
&$X_{SI}$  & 5.2812e+03
     &(2.7451e+03,	9.6715e+03)\\       
$\Sigma$   & 4.2169e-07
	& (4.2102e-07,	4.2238e-07)
&  $X_{RI}$  & 4.3571e+03
 	&(1.3700e+03,	6.7341e+03)\\
$r$  & 7.6724e-02
               & (5.7846e-02,	9.4957e-02)       
&$X_{SR}$  & 2.6449e+04
   &(2.1936e+04,	3.3162e+04)\\
$c$        & 2.1111e-02
              & (1.3201e-02,	2.8164e-02)       
&$X_{IR}$  & 2.6822e+02
     &(1.0254e+01,	1.4385e+03)\\
$\nu$       &  6.0955e-01
                & (4.6341e-01,	7.1377e-01)     
&$X_{RR}$  & 1.4663e+05
     &(1.4151e+05,	1.5063e+05)\\
$v$        & 4.6302e-01
                & (4.6302e-01,	4.6302e-01)       &\\
\hline
\end{tabular}\label{tab:07-08}
\normalsize
\end{table}
\begin{table}
\centering
\caption{Posterior summaries for the year 2008-09}
\begin{tabular}{llllll}
\hline
Parameter  &  MAP    & $95\%$ Credible Intervals      & Parameter&   MAP    & $95\%$ Credible Intervals \\
\hline
$\beta_1$  & 62.0377
               & (56.7042	66.7867)    
&$X_{SS}$  & 2.2821e+06
      &(2.2819e+06,	2.2822e+06)\\
$\beta_2$  & 74.6084
               & (71.1210,	78.4359)   
& $X_{IS}$  & 6.4730e+01
   &(5.6061e+01,	2.9822e+02)\\
$\sigma_1$   & 1.4283
               & (1.1325,	1.8336)       
&$X_{RS}$  & 4.2216e+04
     &(4.2072e+04,	4.2344e+04)\\       
$\sigma_2$   & 1.1053
               & (0.5115,	2.2674)       
&$X_{SI}$  & 3.3911e+02
     &(1.2082e+02,	3.8804e+02)\\        
$\Sigma$   & 3.1394e-06
	& (2.9891e-06,	3.5692e-06)
&  $X_{RI}$  & 9.2281e+01
 	&(4.6576e+01,	2.0449e+02)\\
$r$  & 1.0579e-01
               & (1.0012e-01,	1.0700e-01)       
&$X_{SR}$  & 2.7546e+04
   &(2.7504e+04, 2.7630e+04)\\
$c$        & 3.2293e-02
              & (3.1595e-02,	3.3030e-02)       
&$X_{IR}$  & 2.0122e+02
     &(7.9222e+01,	2.6636e+02)\\
$\nu$       &  2.0002e-01
                & (1.9998e-01,	2.0003e-01)     
&$X_{RR}$  & 1.4748e+05
     &(1.4714e+05,	1.4767e+05)\\
$v$        & 9.9995e-02
                & (9.9994e-02,	9.9996e-02)       &\\
\hline
\vspace{4.5in}
\end{tabular}\label{tab:08-09}
\normalsize
\end{table}
\clearpage
}

\newpage\newpage\newpage\newpage\newpage\newpage\newpage\newpage\newpage
\bibliographystyle{imsart-nameyear} 
\bibliography{BI_references}       